\documentclass[preprint,floatfix,aps,prd,showpacs,footinbib,amsmath,amssymb,amsfonts,superscriptaddress]{revtex4-1}
\usepackage{graphicx}
\usepackage{color}
\usepackage{dsfont}
\usepackage[normalem]{ulem}
\usepackage{hyperref}
\usepackage{mathtools}
\usepackage{mathrsfs}
\usepackage{calrsfs}
\usepackage{comment}
\usepackage[utf8]{inputenc}
\usepackage{varioref}
\usepackage[active]{srcltx}

\expandafter\ifx\csname package@font\endcsname\relax\else
\expandafter\expandafter
\expandafter\usepackage
\expandafter\expandafter
\expandafter{\csname package@font\endcsname}%
\fi
\hypersetup{
	colorlinks=true,       % false: boxed links; true: colored links
	linkcolor=blue,          % color of internal links
	citecolor=blue,        % color of links to bibliography
}
\usepackage[section]{placeins}

\usepackage{fancybox}
\labelformat{figure}{Fig.~#1}

\begin{document}
	\title{Galileon scalar electrodynamics} 
	\author{Ashu Kushwaha} 
	\email{ashu712@iitb.ac.in}
%	\affiliation{Department of Physics, Indian Institute of Technology Bombay, Mumbai 400076, India}
%
	\author{S. Shankaranarayanan}
	\email{shanki@phy.iitb.ac.in}
	\affiliation{Department of Physics, Indian Institute of Technology Bombay, Mumbai 400076, India}

\begin{abstract}
We construct a consistent model of Galileon scalar electrodynamics.
The model satisfies three essential requirements: (1) The action contains higher-order derivative terms, and obey the \emph{Galilean symmetry}, (2) Equations of motion also satisfy Galilean symmetry and contain only up to  second-order derivative terms in the matter fields and, hence do not suffer from instability, and (3) local $U(1)$ gauge invariance is preserved. We show that the non-minimal coupling terms in our model are different from that of the real scalar Galileon models; however, {they match with the Galileon real scalar field action}. We show that the model can lead to an accelerated expansion in the early Universe.  We discuss the implications of the model for cosmological inflation.
\end{abstract}
	\maketitle
\section{Introduction}

The standard relativistic field theories describing physical phenomena contain the second-order time and spatial derivatives. The perturbative approach of these field theories is highly successful in explaining the experiments and observations. However, these field theories have ultraviolet divergences, and higher-derivative terms were introduced in an attempt to remove the ultraviolet divergence~\cite{1948-Podolsky.Schwed-RMP,1950-Thirring-PR,1950-Pais.Uhlenbeck-PR}. 

The higher derivative field theories are known to have Ostrogradsky instability~\cite{2015-Woodard-arXiv}. This is because Hamiltonian contains a term proportional to the momenta, thus leading to an unbounded Hamiltonian from below~\cite{1990-Simon-PRD}. The instability itself is not a concern if the energy is an integral of motion or the higher derivative field (say, $\pi$) does not interact with any other field (say, $\phi$) whose Hamiltonian is bounded from below. The interaction between $\pi$ and $\phi$ will pump out energy from $\pi$, leading to a runaway situation. Hence, these negative energy states can be traded by negative norm states (or ghosts), leading to non-unitary theories~\cite{2002-Hawking.Hertog-PRD} and, therefore, unsuitable to describe physical phenomena.

Nonetheless, higher-derivative theories have some salient features that
make them indispensable to understand the high-energy behavior of a theory~\cite{1983-Barth.Christensen-PRD}. Specifically, the divergence structure of quantum fields is expected to improve when higher-derivative terms are taken into account~\cite{1983-Barth.Christensen-PRD,2002-Hawking.Hertog-PRD}. This also leads to improved convergence of Feynman diagrams. Although Einstein's gravity is not renormalizable, conformal gravity is~\cite{1977-Stelle-PRD,1982-Fradkin.Tseytlin-NPB}.

There is a resurgence of interest in higher-derivative theories primarily from the modifications of gravity in the short and long distances~\cite{2010-Sotiriou.Faraoni-RMP,2011-Capozziello.DeLaurentis-PRep,2011-Nojiri.Odintsov-PRep,2012-Clifton.etal-PRep,2017-Nojiri.etal-PRep,Ishak:2018his}. 
It is now known that higher derivative gravity theories like $f(R)$ are degenerate, and do not suffer from Ostrogradsky instability~\cite{2007-Woodard-Proc}. A consistent scheme has been implemented where the Ostrogradsky instability is suppressed by requiring that the equations of motion are second order~\cite{1974-Horndeski-IJTP}. Starting from the following general action:
\begin{equation}
L=L\left(g_{i j} ; g_{i j, i_{1}} ; \ldots ; g_{i j, i_{1} \cdots i_{p}} ; \phi ; \phi ; \phi,_{i_{1}} ; \ldots ; \phi_{, i_{1}} \ldots i_{q}\right) \, ,
\end{equation}
Horndeski showed that it is possible to obtain a second-order equation for $\phi$ even if the action contains higher-order derivatives of the scalar field~\cite{1974-Horndeski-IJTP}. It has recently shown that the field equations in a subclass of Horndeski models have Galilean shift symmetry in flat space, i.e., 
\begin{equation}
\label{eq:GalileanShift}
\phi \rightarrow \phi + a_{\mu}x^{\mu} + b \, ,
\end{equation}
where $a_{\mu}$ and $b$ are infinitesimal four-vector (shift parameter) and a constant scalar respectively~\cite{2009-Deffayet.etal4-PRD,2014-Deffayet.etal-JHEP}. Galileon scalar fields are the most general non-canonical and non-minimally coupled single-field model, which yields second-order equations.

These theories, while naively look like \emph{higher-derivative} theories, are healthy non-higher-derivative theories; their equations of motion are second order in time derivatives and do not suffer from the instability~\cite{2009-Deffayet.etal4-PRD,2014-Deffayet.etal-JHEP}. This is achieved by the addition of structure in the Lagrangian --- usually by a subtle cancellation of higher derivative terms in the equations of motion, much like Lanczos-Lovelock theories of gravity~\cite{1971-Lovelock-JMP,1972-Lovelock-JMP,1986-Zumino-PRep,2013-Padmanabhan.Kothawala-PRep}. 
Like Galileon scalars, these theories of gravity are exceptional in that the resulting equations of motion are no more than the second order. They are also free of ghosts when expanded about the flat space-time.

Recently, a vector Galileon field was constructed in a curved space-time~\cite{2017-Nandi.Shankaranarayanan-JCAP}. It was shown that the electromagnetic action breaks conformal invariance and has the following three properties: model is described by vector potential $A_{\mu}$ and its derivatives, Gauge invariance is preserved, and equations of motion are linear in second derivatives of the vector potential. This is an essential result as earlier it was proven that such an action could not be constructed in flat space-time~\cite{2014-Deffayet.etal-JHEP}.

While the scalar and vector Galileon fields have been investigated, there has been \emph{no study} on the interaction between the scalar and vector Galileons \emph{preserving gauge-invariance.} (In Ref. \cite{2018-Heisenberg.etal-PRD}, Heisenberg et al have studied Scalar-Vector Galileon models that do not preserve gauge invariance.) In this work, we bridge this gap by constructing a higher derivative action of interacting fields (complex scalar Galileons interacting with a vector field), which do not lead to ghosts. We refer to this model as \emph{Galileon scalar electrodynamics}. We construct a complex scalar Galileon action in flat space-time by demanding the following conditions: action must satisfy the Galileon symmetry, equations of motion must be second-order, gauge invariance is preserved. Naturally, the new action has the standard scalar electrodynamics with new terms that contribute at high-energies. We apply this model in the early Universe. 
%In the Coulomb gauge of the Vector field, we obtain the non-zero contributions from the new terms. In the flat FLRW background, we show that the model leads to inflation and is driven by the real scalar field. 

The rest of the manuscript is organized as follows: 
In Sec. \eqref{sec:ComplexGalileon}, we explicitly construct the 
Galileon scalar electrodynamics in flat space-time. We show that 
the action is invariant under the local gauge invariance in flat space-time. In Sec. \eqref{Min}, we extend the analysis to curved space-time. We show that the minimal coupling of the matter and gravity leads to higher-derivative terms in the equations of motion. We then include non-minimal terms to the action that will lead to subtle cancellation of terms and, hence, lead to second-order equations of motion. We then apply the model to the early Universe in Sec. \eqref{sec:energy_pressure} and show that the model can lead to inflation. In Sec. \eqref{sec:conc}, we conclude by briefly discussing the importance of these results. Appendices (A-F) contain the details of the calculations in the main text. 

In this work, we use (+,-,-,-) metric signature and natural units $\hbar = c = 1/(4\pi \epsilon_0) = 1$. We set $8\pi G = M_P^{-2}$.  
The real Galileon scalar field is denoted by $\varphi$, complex Galileon scalar field is denoted by $\pi$ and  $\Box = \nabla_{\mu}\nabla^{\mu}$. An overdot denotes the derivative for cosmic time.

\section{Galileon Scalar Electrodynamics in flat space-time}
\label{sec:ComplexGalileon}

In this section, we obtain the higher derivative complex scalar field action coupled to the electromagnetic field in flat space-time. In the next subsection, we list the real scalar Lagrangians obtained by Nicolis et al.~\cite{2009-Nicolis-PRD} that lead to \emph{healthy} non-higher-derivative theories. In Sec. \eqref{sec:CSG}, we obtain the complex scalar Lagrangian that leads to non-higher-derivative theories. In Sec. \eqref{sec:GaugeFlat}, we couple the Galileon complex scalar with the electromagnetic field $A_{\mu}$. 

\subsection{Real scalar Galileon}
\label{sec:ScalarGalileon}

As mentioned earlier, the action of the real scalar Galileons ($\varphi$) is invariant under the \emph{Galilean transformation} \eqref{eq:GalileanShift}. In 4-D Minkowski space-time, Nicolis et al.~\cite{2009-Nicolis-PRD} have shown that the following five Lagrangians are invariant under the Galilean transformation:
\begin{subequations}
\begin{align} \label{eq:ScalarL2}
    & \mathcal{L}_1^{\prime} = \varphi \qquad 
    \mathcal{L}_2^{\prime} = -\frac{1}{2}\partial_{\mu}\varphi\partial^{\mu} \varphi\\
    \mathcal{L}_3^{\prime} &= -\frac{1}{2}\Box \varphi\partial_{\mu}\varphi\partial^{\mu} \varphi\\
    \label{eq:ScalarL4}
    \mathcal{L}_4^{\prime} &= -\frac{1}{4}[ (\Box \varphi)^2 \partial_{\mu}\varphi\partial^{\mu} \varphi -2\Box \varphi\partial_{\mu}\varphi\partial^{\mu}\partial_{\nu}\varphi\partial^{\nu}\varphi- \partial_{\mu}\partial_{\nu}\varphi \partial^{\mu}\partial^{\nu}\varphi\partial_{\alpha}\varphi\partial^{\alpha} \varphi +2\partial_{\mu}\varphi \partial^{\mu}\partial_{\nu}\varphi\partial^{\nu}\partial_{\alpha}\varphi
    \partial^{\alpha}\varphi ]\\
    \mathcal{L}_5^{\prime} &= -\frac{1}{5}[ (\Box \varphi)^3 \partial_{\mu}\varphi\partial^{\mu} \varphi -3(\Box \varphi)^2 \partial_{\mu}\varphi \partial^{\mu}\partial_{\nu}\varphi\partial^{\nu} \varphi - 3\Box \varphi\partial_{\mu}\partial^{\nu}\varphi\partial^{\mu}\partial_{\nu}\varphi\partial_{\alpha}\varphi \partial^{\alpha} \varphi \nonumber\\
    &{}{}\hspace{0.5cm}+ 6\Box \varphi\partial_{\mu}\varphi\partial^{\mu}\partial_{\nu}\varphi \partial^{\nu}\partial_{\alpha}\varphi\partial^{\alpha} \varphi 
+2\partial_{\mu}\partial^{\nu}\varphi\partial_{\nu}\partial^{\alpha}\varphi\partial_{\alpha}\partial^{\mu}\varphi\partial_{\lambda}\varphi \partial^{\lambda} \varphi + 3\partial_{\mu}\partial^{\nu}\varphi\partial_{\nu}\partial^{\mu}\varphi\partial_{\alpha}\varphi \partial^{\alpha}\partial_{\lambda}\varphi\partial^{\lambda} \varphi \nonumber\\
&{}{}\hspace{0.5cm}-6\partial_{\mu}\varphi \partial^{\mu}\partial_{\nu}\varphi\partial^{\nu}\partial_{\alpha}\varphi\partial^{\alpha}\partial_{\lambda}\varphi\partial^{\lambda} \varphi ]
    \end{align}
    \end{subequations}
 We want to emphasize that the above Lagrangians are the linear combinations of Lorentz invariant terms which are added (up to total derivative term) in such a way that the action is Galilean invariant. The equations of motion are second order. In the next subsection, we use these five Lagrangians to construct the complex scalar Galileon action.
 
    \subsection{Complex scalar Galileon}
    \label{sec:CSG}
    
We aim to write down the Lagrangian that satisfies the following three conditions: First, it must satisfy shift symmetry (\ref{eq:GalileanShift}) on $\pi$ and the complex conjugate ($\pi^*$). Second, it must be Lorentz invariant. Third, the Lagrangian must be real. It is easy to see that the Lagrangians with the odd number of $\pi$'s lead to complex action and, hence, are discarded. Having $\mathcal{L}_2^{\prime}$ and $\mathcal{L}_4^{\prime}$ in hand, we start with $\mathcal{L}_2^{\prime}$. The complex scalar Galileon Lagrangian for this case is straightforward:
    \begin{align}\label{L2a}
    \mathcal{L}_2^{\prime} = \frac{1}{2}\partial_{\mu}\pi \partial^{\mu}\pi \implies {\mathcal{L}_2} = \frac{1}{2}\partial_{\mu}\pi \partial^{\mu}\pi^*
    \end{align}
 where $\pi^*$ is the complex conjugate. The above action is invariant 
 under the Galilean shift symmetry (\ref{eq:GalileanShift}) on $\pi$ and $\pi^*$.  In order to avoid confusion, we have used unprimed for complex scalar Lagrangian to distinguish it from the scalar Galileon Lagrangian (primed). 
 
Next, we consider the Lorentz invariant terms of $\mathcal{L}_4^{\prime}$ (with two fields $\pi$ and $\pi^*$). By symmetrizing each term in 
the RHS of \eqref{eq:ScalarL4} and ignoring the double-counting, 
we write $\mathcal{L}_4$ as a linear combination of these invariants, i.e.,
    \begin{align}\label{L4}
    \mathcal{L}_4 &= A_1\, (\, (\Box\pi)^2 \partial_{\alpha}\pi^* \partial^{\alpha}\pi^* + \rm{c.c.} ) + A_2 \,\,\Box \pi^* \Box \pi \partial_{\alpha}\pi^* \partial^{\alpha}\pi   
    +  B_1(\, \Box \pi\partial_{\nu}\pi \partial^{\nu} \partial^{\alpha}\pi^* \partial_{\alpha}\pi^*+ \rm{c.c.}) \nonumber \\ &{}+ B_2\,(\, \Box \pi\partial_{\nu}\pi^* \partial^{\nu} \partial^{\alpha}\pi \partial_{\alpha}\pi^* + \rm{c.c.})  + C_1\,(\, \partial_{\mu}\partial_{\nu}\pi \partial^{\mu}\partial^{\nu}\pi \partial^{\alpha}\pi^* \partial_{\alpha}\pi^* + \rm{c.c.}) + C_2\,\, \partial_{\mu}\partial_{\nu}\pi^*\partial^{\mu}\partial^{\nu}\pi \partial^{\alpha}\pi^* \partial_{\alpha}\pi \nonumber \\ &{}
    + D_1 \,(\partial_{\mu}\pi^*\partial^{\mu}\partial^{\nu}\pi \partial_{\nu}\partial_{\alpha}\pi \partial^{\alpha}\pi^* + \rm{c.c.}) + 
    D_2 \,\,\partial_{\mu}\pi^*\partial^{\mu}\partial^{\nu}\pi^* \partial_{\nu}\partial_{\alpha}\pi \partial^{\alpha}\pi +  D_3 \,\,\partial_{\mu}\pi^*\partial^{\mu}\partial^{\nu}\pi \partial_{\nu}\partial_{\alpha}\pi^* \partial^{\alpha}\pi
    \end{align}
where $A_i, B_i, C_i, D_i, D_3 \, (i = 1, 2)$ are unknown complex constants, c.c. in the parentheses denote the complex conjugate part (hence total part contributing to real) of the corresponding Lorentz invariant term and rest of the terms are real. Hence, the above  Lagrangian is a real scalar.

However, for any arbitrary constants, the Lagrangian will not be Galilean invariant. Demanding the Galilean invariance of the Lagrangian \eqref{L4} (up to some total derivative) leads to the constraints on the constants. Appendix \eqref{Generic} contains detailed calculations 
where we show that $\mathcal{L}_4 $ depends only on 
two arbitrary coefficients $A_1$ and $A_2$ and the resultant equations of motion match with Nicolis et al.~\cite{2009-Nicolis-PRD}. In Appendix \eqref{Generic}  we also explicitly show that the equations of motion for this Lagrangian is the same for any value of $A_1$ and $A_2$. This is because the action $\mathcal{L}_4 $ is invariant under the shift symmetry \eqref{eq:GalileanShift} for any value of these arbitrary constants. For simplicity, we set $A_1 = 0$. Setting $A_1 = 0$ and $A_2 = \omega \, \lambda^{-6}$ in Eq.~(\ref{LCGG}) leads to:
\begin{align}\label{GIL}
\mathcal{L}_4 &= \frac{\omega}{2 \lambda_{\pi}^6} \left[ 2\, \Box \pi^* \,\Box \pi \,\partial_{\alpha}\pi \partial^{\alpha}\pi^* 
- 2\, \partial_{\mu}\partial_{\nu}\pi^* \partial^{\mu}\partial^{\nu}\pi \partial_{\alpha} \pi\partial^{\alpha}\pi^* \right. \nonumber \\
&+ \left. (\,\Box \pi \,\partial_{\nu}\pi^* \partial^{\nu}\partial^{\alpha}\pi \partial_{\alpha}\pi^* 
- \partial^{\mu}\partial^{\nu}\pi \partial_{\nu}\partial_{\alpha}\pi\,\partial_{\mu}\pi^* \partial^{\alpha}\pi^* +\rm{c.c.}) \right]
\end{align}
Note that $A_2$ has a dimension $[L]^{6}$, thus $\lambda$ has dimensions of inverse length and $\omega$ can take $+1$ or $-1$. 
The value of $\omega$ will be fixed in Sec. \eqref{sec:coeffiFixing} by demanding that the energy density of the field is always positive. In the next section, we will show that the generalization of the above Lagrangian in curved space-time matches with Deffayet et al. for some suitable value of the coefficient $\lambda_{\pi}$~\cite{Deffayet2009}. The Galilean invariant action corresponding to the above Lagrangian is given by
\begin{align}\label{action4}
{S}_{4} &= \frac{\omega}{2 \lambda_{\pi}^6}\int d^4x\,\,[\,2\, \Box \pi^* \,\Box \pi \,\partial_{\alpha}\pi \partial^{\alpha}\pi^* - 2\, \partial_{\mu}\partial_{\nu}\pi^* \partial^{\mu}\partial^{\nu}\pi \partial_{\alpha} \pi\partial^{\alpha}\pi^*   \nonumber\\&{}\hspace{2cm} +  (\,\Box \pi \,\partial_{\nu}\pi^* \partial^{\nu}\partial^{\alpha}\pi \partial_{\alpha}\pi^* 
- \partial^{\mu}\partial^{\nu}\pi \partial_{\nu}\partial_{\alpha}\pi\,\partial_{\mu}\pi^* \partial^{\alpha}\pi^* +\rm{c.c.}) ] 
\end{align}
Equations of motion for $\pi$ corresponding to the above action is:
\begin{align}\label{EOM}
\mathcal{E}_{4} = \frac{\omega}{2 \lambda_{\pi}^6}\left[ -(\Box \pi^*)^2\,\Box \pi + \Box \pi\, \partial_{\mu}\partial_{\nu}\pi^* \,\partial^{\mu} \partial^{ \nu}\pi^* - 2\,\partial_{\mu}\partial_{\nu}\pi^*\, \partial^{\nu}\partial^{\alpha}\pi\,\partial^{\mu}\partial_{\alpha}\pi^* + 2\,\Box \pi^*\, \partial_{\mu}\partial_{\nu}\pi^*\, \partial^{\mu}\partial^{\nu}\pi\,\,\right]
\end{align}
The equations of motion contain derivatives up to second order and, hence, does not lead to extra degrees of freedom. The complex scalar Galileon action in 4D Minkowski space-time is given by
\begin{align}\label{action_CSG}
{S}_{\rm{flat}} &= S_{2} + S_{\rm 4} \, ,
\end{align}
where $S_{\rm 4}$ is given by \eqref{action4} and 
\begin{equation}
\label{eq:CanonicalScalar}
S_{2} = \frac{1}{2}\int d^4x \,\,\partial_{\mu}\pi\, 
\partial^{\mu}\pi^*  \, .
\end{equation}
Thus, we have constructed a complex Galileon field action. In the low-energy limit, the Galileon term will not be significant. However, at high-energies $S_4$ plays a crucial role in the dynamics. 
We will discuss the implications of this in Sec. \eqref{sec:energy_pressure}.

\subsection{Coupling to the electromagnetic field}
\label{sec:GaugeFlat}

The action \eqref{action_CSG} is invariant under the global transformation, i.e. $\pi \to \pi e^{-i\, e \, \theta}$, where $\theta$ is a constant parameter and $e$ is the electric charge. However, the action is not invariant if the parameter $\theta$ is space-time dependent or the local $U(1)$ gauge transformation.  In order 
for the action to be invariant under the transformation: 
$\pi \rightarrow \pi e^{-ie\theta(x)}$ and $\pi^* \rightarrow \pi e^{ie\theta(x)}$, we need to replace the partial derivatives as
\begin{align}\label{eq:EMcovariant}
\partial_{\mu} \rightarrow {D}_{\mu} \equiv \partial_{\mu} + ie A_{\mu} \hspace{0.5cm} \text{and} \hspace{0.5cm}
\partial_{\mu}\partial^{\nu} \rightarrow {D}_{\mu}{D}^{\nu}= (\partial_{\mu} + ie A_{\mu} )(\partial^{\nu} + ie A^{\nu})
\end{align}
where $A_{\mu}$ is the electromagnetic field vector. Under the following gauge transformations 
\begin{align}\label{eq:GaugeTransformation}
\pi \rightarrow \pi e^{-ie\theta(x)}\,;\hspace{.5cm}\pi^* \rightarrow \pi e^{ie\theta(x)}\,; \hspace{.5cm}
A_{\mu} \rightarrow A_{\mu} + \partial_{\mu}\theta
\end{align}
the action (\ref{action_CSG}) is invariant. Appendix \eqref{GaugeAppend} contains the details of the calculation. 
The complete Galileon scalar electrodynamics action in flat space-time is given by:
\begin{align}\label{action_full}
{S}_{\rm{flat}}^{\rm G} &= \frac{1}{2}\int d^4x\,\, {D}_{\mu}\pi\, {D}^{\mu}\pi^*+\frac{\omega}{2 \lambda_{\pi}^6}\int d^4x\,\,[\,2\, {D}_{\mu}{D}^{\mu} \pi^*\, {D}_{\nu}{D}^{\nu} \pi \,{D}_{\alpha}\pi \,{D}^{\alpha}\pi^* - 2 \,{D}_{\mu}{D}_{\nu}\pi^*\,{D}^{\mu}{D}^{\nu}\pi\, {D}_{\alpha} \pi\, {D}^{\alpha}\pi^*\nonumber\\&{}\hspace{0.5cm} + ( {D}_{\mu}{D}^{\mu} \pi \,{D}_{\nu}\pi^*\, {D}^{\nu}{D}^{\alpha}\pi \,{D}_{\alpha}\pi^*   
 - {D}^{\mu}{D}^{\nu}\pi\, {D}_{\nu}{D}_{\alpha}\pi\,{D}_{\mu}\pi^*\,{D}^{\alpha}\pi^* + \rm{c.c.}) ] -\frac{1}{4}\int d^4 x \,\,F_{\mu\nu}\,F^{\mu\nu}
\end{align}
where $F_{\mu\nu} = \partial_{\mu}A_{\nu} - \partial_{\nu}A_{\mu}$ is the electromagnetic field tensor. Unlike scalar Galileons, there exists a 
no-go theorem that states that, higher derivative vector Galileons cannot be constructed in flat space-time~\cite{2014-Deffayet.etal-JHEP,2017-Nandi.Shankaranarayanan-JCAP}. Thus, for the Galileon scalar electrodynamics in the flat space-time, the non-linear part of the electromagnetic fields appear through the gauge coupling $D_{\mu}$. It is important to note that the action (\ref{action_full}), in addition to U(1) gauge invariance, also satisfies the Galilean symmetry (for $\pi$ and $\pi^*$), and hence, we refer to  this action as \emph{Galileon Scalar Electrodynamics}. In Sec. \eqref{sec:energy_pressure}, we show that the Galileon term leads to interesting features in inflationary dynamics.

\section{Galileon Scalar Electrodynamics in curved space-time}\label{Min}

One of the critical features of the action of the Galileon field 
\eqref{action_CSG} is that it contains second derivatives of the field. 
Assuming a minimal coupling of the matter and gravity leads to $\partial_{\mu} \to \nabla_{\mu}$. While the partial derivatives $\partial_{\mu}\partial_{\nu}$ commute, this is not the case for covariant derivatives. We need to take into account the commutation properties of the covariant derivatives. The procedure we will adopt is similar to that of Deffayat et al~\cite{Deffayet2009}. However, due to complex scalar fields, there are some differences in the final expression.

\subsection{Coupling to gravity}
\label{sec:CurvedST}

For the minimal coupling, Galilean symmetry is preserved for $\mathcal{L}_2$ defined in \eqref{L2a}. However, the 
Galilean symmetry is broken explicitly for the fourth order Lagrangian (\ref{GIL}). Assuming a minimal coupling of the complex field with gravity, the fourth order action (\ref{action4}) becomes:
\begin{align}\label{action_m}
{S}_4^{\rm{min}} &= \frac{\omega}{2 \lambda_{\pi}^6}\int d^4x\,\,\sqrt{-g}\,\,[\,\,2\, \Box \pi^*\, \Box \pi \,\nabla_{\alpha}\pi\, \nabla^{\alpha}\pi^* - 2 \,\nabla_{\mu}\nabla_{\nu}\pi^*\, \nabla^{\mu}\nabla^{\nu}\pi\, \nabla_{\alpha} \pi\, \nabla^{\alpha}\pi^*\nonumber\\&{}\hspace{0.5cm} + (\,\, \Box \pi\, \nabla_{\nu}\pi^* \, \nabla^{\nu}\nabla^{\alpha}\pi \,\nabla_{\alpha}\pi^* 
  - \nabla^{\mu}\nabla^{\nu}\pi\, \nabla_{\nu}\nabla_{\alpha}\pi\,\nabla_{\mu}\pi^*\, \nabla^{\alpha}\pi^*  + \rm{c.c.} )\,\,\,] 
\end{align}
here $\nabla_{\alpha}$ denotes the covariant derivative with respect to the metric $g_{\mu\nu}$ and $\Box = \nabla_{\mu}\nabla^{\mu}$. Varying the action (\ref{action_m}) with respect to $\pi$ yields the equation of motion of $\pi$. Using the commutation properties of covariant derivatives as given in Appendix \eqref{app:C}, we obtain the 
following equation of motion:
\begin{align}
\label{eq:EOM4-Curved}
\mathcal{E}_4^{\rm{min}} = \frac{\omega}{2 \lambda_{\pi}^6}[\,-2\,\nabla_{\alpha}\pi\, \nabla^{\alpha}\pi^*\,\nabla^{\nu}\nabla^{\mu}\pi^*\,R_{\mu\nu} -\nabla_{\alpha}\pi\, \nabla^{\alpha}\pi^*\,\nabla^{\mu}\pi^*\,\nabla_{\mu}R - \nabla^{\mu}\pi^*\,\nabla^{\alpha}\pi^*\,\nabla^{\rho}\pi\,\nabla_{\rho}R_{\alpha\mu}
 \nonumber\\ - \nabla^{\mu}\pi^*\,\nabla^{\alpha}\pi^*\,\nabla_{\mu}\nabla^{\nu}\pi\, R_{\nu\alpha} +2\,\nabla^{\mu}\pi^*\,\nabla^{\alpha}\pi^*\,\nabla^{\nu}\nabla^{\rho}\pi\, R_{\rho\mu\alpha\nu} -3\,\nabla^{\alpha}\nabla^{\nu}\pi\, \nabla_{\alpha}\pi^* \,\nabla^{\mu}\pi^*\, R_{\mu\nu}\nonumber\\ -2\,\nabla^{\mu}\pi^*\,\nabla_{\alpha}\pi\,\nabla^{\nu}\nabla^{\alpha}\pi^*\,R_{\mu\nu} + \nabla^{\nu}\pi^*\,\nabla^{\alpha}\pi^*\,\Box\pi\, R_{\nu\alpha} -2\,\nabla^{\nu}\nabla^{\alpha}\pi^*\,\nabla_{\alpha}\pi^*\,\nabla^{\mu}\pi\, R_{\mu\nu}\nonumber\\ +2\,\Box\pi^*\, \nabla_{\nu}\nabla_{\alpha}\pi\,\nabla^{\nu}\nabla^{\alpha}\pi^* -(\Box\pi^*)^2\, \Box\pi - 2\,\nabla_{\mu}\nabla_{\nu}\pi^*\, \nabla^{\mu}\nabla^{\alpha}\pi^*\nabla^{\nu}\nabla_{\alpha}\pi +\Box\pi\, \nabla^{\mu}\nabla^{\nu}\pi^* \,\nabla_{\mu}\nabla_{\nu}\pi^*\,\,]
\end{align}
where $R_{\rho\mu\alpha\nu}, R_{\mu\nu}$ and $R$ are Riemann tensor, Ricci tensor and Ricci scalar respectively. One can immediately notice that the second and third terms in the RHS of the above
contain third order derivative terms. Thus, the minimal coupling of the matter with gravity lead to the higher-derivative equations of motion. 
As mentioned in the introduction, it is possible to cancel these higher derivative terms adding suitable terms in the action~\cite{Deffayet2009,2017-Nandi.Shankaranarayanan-JCAP}. The following non-minimal action 
\begin{align}\label{non-min}
{S}_4^{\rm{nm}}&= -\frac{\omega}{4 \lambda_{\pi}^6}\int d^4x\,\,\sqrt{-g}\,\, \nabla_{\alpha}\pi\nabla^{\alpha}\pi^* \nabla_{\mu}\pi \nabla_{\nu}\pi^*g^{\mu\nu}R \nonumber\\&{} \nonumber\\ &{}\hspace{0.5cm}  -\frac{\omega}{4 \lambda_{\pi}^6}\int d^4x\,\,\sqrt{-g}\,\, \left[\nabla_{\alpha}\pi\nabla^{\alpha}\pi \nabla_{\mu}\pi^*\nabla_{\nu}\pi^* \left( R^{\mu\nu} - \frac{1}{4}g^{\mu\nu}R \right) +\rm{c.c.} \right]
\end{align}
can remove the higher-derivative terms that appear in Eq. \eqref{eq:EOM4-Curved}. See Appendices \eqref{app:C} and \eqref{sec:A1_terms}, for details. 
If we add the above non minimal action in the action (\ref{action_m}) then it will cancel all the higher order derivative terms in the EOM. So varying the action
${S}_4^{\rm{min}} + {S}^{\rm{nm}}_4$ with respect to $\pi$ gives the equation of motion  (see appendix Eq. \ref{eq:S4eomCurved}).
%
%Hence equation of motion for the total action 
%\begin{align}\label{S'}
%{S}^{\prime}_{\rm{CGG}} = {S}_{\rm{CGG}} + {S}_{\rm{non\; min}}
%\end{align}
%will have only second order terms in the Galileon fields and metric, which is 
%\begin{align}\label{Eprime}
%\mathcal{E}^{\prime} &= \frac{A_2}{2}\,\,[\,\,2\,\Box\pi^*\, \nabla_{\nu}\nabla_{\alpha}\pi\,\nabla^{\nu}\nabla^{\alpha}\pi^* -(\,\Box\pi^*)^2\, \Box\pi +\Box\pi\, \nabla^{\mu}\nabla^{\nu}\pi^*\, \nabla_{\mu}\nabla_{\nu}\pi^* - 2\,\nabla_{\mu}\nabla_{\nu}\pi^*\, \nabla^{\mu}\nabla^{\alpha}\pi^*\,\nabla^{\nu}\nabla_{\alpha}\pi \nonumber\\ &{} \hspace{1cm} + \Box\pi^*\,\nabla_{\mu}\pi^*\,\nabla^{\mu}\pi\, R -\frac{1}{2}\Box\pi\,\nabla_{\alpha}\pi^*\,\nabla^{\alpha}\pi^* \,R + \nabla_{\alpha}\pi^*\,\nabla_{\mu}\pi^*\,\nabla^{\mu}\nabla^{\alpha}\pi \,R +2\,\Box\pi\, \nabla^{\nu}\pi^*\,\nabla^{\mu}\pi^* \,R_{\nu\mu} \nonumber\\ &{} \hspace{1cm} +\nabla_{\alpha}\pi^*\, \nabla^{\alpha}\pi^*\,\nabla^{\nu}\nabla^{\mu}\pi \,R_{\mu\nu}  - 2\,\nabla_{\alpha}\pi\, \nabla^{\alpha}\pi^*\,\nabla^{\nu}\nabla^{\mu}\pi^*\,R_{\mu\nu} - \nabla^{\mu}\pi^*\,\nabla^{\alpha}\pi^*\,\nabla_{\mu}\nabla^{\nu}\pi\, R_{\nu\alpha} \nonumber\\ &{} \hspace{1cm}+ 2\, \nabla^{\mu}\pi^*\,\nabla^{\alpha}\pi^*\,\nabla^{\nu}\nabla^{\rho}\pi\, R_{\rho\mu\alpha\nu} -3\,\nabla^{\alpha}\nabla^{\nu}\pi\, \nabla_{\alpha}\pi^*\, \nabla^{\mu}\pi^*\,R_{\mu\nu} ]
%\end{align}
%
Hence, the complete Galileon Complex scalar action in an arbitrary curved space-time is given by
\begin{align}\label{action_curved}
{S}_{\rm{Curved}} &= S_2  + S_4^{\rm{min}} + S_4^{\rm{nm}} \, .
%
%\frac{1}{2}\int d^4x\, \sqrt{-g}\,\,\nabla_{\mu}\pi\nabla^{\mu}\pi^* +\frac{A_2}{2}\int d^4x\,\,\sqrt{-g}\,\,[\,\,2\, \Box \pi^*\, \Box \pi \,\nabla_{\alpha}\pi\, \nabla^{\alpha}\pi^* - 2 \,\nabla_{\mu}\nabla_{\nu}\pi^*\, \nabla^{\mu}\nabla^{\nu}\pi\, \nabla_{\alpha} \pi\, \nabla^{\alpha}\pi^*\nonumber\\&{}\hspace{0.5cm} + (\,\, \Box \pi\, \nabla_{\nu}\pi^* \, \nabla^{\nu}\nabla^{\alpha}\pi \,\nabla_{\alpha}\pi^* 
%- \nabla^{\mu}\nabla^{\nu}\pi\, \nabla_{\nu}\nabla_{\alpha}\pi\,\nabla_{\mu}\pi^*\, \nabla^{\alpha}\pi^*  + \rm{c.c.} )\,\,\,] \nonumber\\&{}\hspace{0.5cm}
%-\frac{A_2}{4}\int d^4x\,\,\sqrt{-g}\,\, \nabla_{\alpha}\pi\nabla^{\alpha}\pi^* \nabla_{\mu}\pi \nabla_{\nu}\pi^*g^{\mu\nu}R   -\frac{A_2}{4}\int d^4x\,\,\sqrt{-g}\,\, \left[\nabla_{\alpha}\pi\nabla^{\alpha}\pi \nabla_{\mu}\pi^*\nabla_{\nu}\pi^* \left( R^{\mu\nu} - \frac{1}{4}g^{\mu\nu}R \right) + \rm{c.c.} \right]
\end{align}
This is one of the key result regarding which we would like to stress the following: First, the non-minimal coupling terms in \eqref{non-min} is different for the complex scalar as compared to the non-minimal terms that arises for the real scalar field~\cite{Deffayet2009}. Second, as expected, the non-minimal coupling terms vanish and the above action matches with flat space time action \eqref{action_CSG}. Third, in the limit of $\pi = \pi^*$, $S_4^{\rm min} $ and $S_4^{nm}$ reduce to:
\begin{subequations}
\label{eq:S4pi}
\begin{align}
\left. {S}_4^{min} \right|_{\pi = \pi^*}  &= \frac{\omega}{\lambda_{\pi}^6}\int d^4x\,\,\sqrt{-g}\,\, \left[\, (\Box\pi)^2 \,\nabla_{\alpha}\pi\, \nabla^{\alpha}\pi + \Box \pi\, \nabla_{\nu}\pi \, \nabla^{\nu}\nabla^{\alpha}\pi \,\nabla_{\alpha}\pi  \right.
\nonumber\\
\label{eq:S4minpi}
& ~~ - \left. \,\nabla_{\mu}\nabla_{\nu}\pi\, \nabla^{\mu}\nabla^{\nu}\pi\, \nabla_{\alpha} \pi\, \nabla^{\alpha}\pi 
 - \nabla^{\mu}\nabla^{\nu}\pi\, \nabla_{\nu}\nabla_{\alpha}\pi\,\nabla_{\mu}\pi\, \nabla^{\alpha}\pi\, \right] \\
\left. {S}_4^{nm} \right|_{\pi = \pi^*}  &=
- \frac{\omega}{4 \, \lambda_{\pi}^6}\int d^4x\,\,\sqrt{-g}\,\, \nabla_{\alpha}\pi\nabla^{\alpha}\pi \nabla_{\mu}\pi \nabla_{\nu}\pi g^{\mu\nu}R \nonumber\\
\label{eq:S4nmpi}
&{}   - \frac{\omega}{2 \lambda_{\pi}^6} \int d^4x\,\,\sqrt{-g}\,\,\nabla_{\alpha}\pi \nabla^{\alpha}\pi \nabla_{\mu}\pi\nabla_{\nu}\pi\,\,\left( R^{\mu\nu} - \frac{1}{4}g^{\mu\nu}R \right)
\end{align}
\end{subequations}
Although the above action looks different compared to that of  Deffayet et al \cite{Deffayet2009}, it is possible to show that the two actions are related by a boundary term. In Appendix \eqref{sec:consistencyRealGal}, we have explicitly shown that the above action for real scalar field is identical to the action used in Ref.~\cite{Deffayet2009}. Hence, the equations of motion from the above action matches with the equations of motion derived in Ref. \cite{Deffayet2009}.

\subsection{Coupling to the electromagnetic field}
\label{G}

Like in the flat space-time, the action \eqref{action_curved} is 
invariant under the global transformation, i.e. $\pi \to \pi e^{-i\, e \, \theta}$, where $\theta$ is a constant parameter and $e$ is the electric charge. In order for the action to be invariant under the 
local gauge transformation, we can replace $\nabla_{\mu} \to \mathcal{D}_{\mu} \equiv \nabla_{\mu} + ieA_{\mu}$ in action \eqref{action_curved}. Beside this, the electromagnetic field will have additional Galileon terms that vanish in the flat space-time~\cite{2017-Nandi.Shankaranarayanan-JCAP}. The complete Galileon scalar electrodynamics action in curved space time is given by
\begin{align}\label{action_full_curved}
{S}^{\rm{G}}_{Curved} &=  \frac{1}{2}\int d^4x\sqrt{-g} \mathcal{D}_{\mu}\pi \mathcal{D}^{\mu}\pi^*+\frac{\omega}{2 \lambda_{\pi}^6}\int d^4x \sqrt{-g}[\,\,2 \mathcal{D}_{\mu}\mathcal{D}^{\mu} \pi^* \mathcal{D}_{\nu}\mathcal{D}^{\nu} \pi \mathcal{D}_{\alpha}\pi \mathcal{D}^{\alpha}\pi^*\nonumber\\&{}- 2 \mathcal{D}_{\mu}\mathcal{D}_{\nu}\pi^* \mathcal{D}^{\mu}\mathcal{D}^{\nu}\pi \mathcal{D}_{\alpha} \pi \mathcal{D}^{\alpha}\pi^* + (\, \mathcal{D}_{\mu}\mathcal{D}^{\mu} \pi \mathcal{D}_{\nu}\pi^*  \mathcal{D}^{\nu}\mathcal{D}^{\alpha}\pi \mathcal{D}_{\alpha}\pi^*  
 - \mathcal{D}^{\mu}\mathcal{D}^{\nu}\pi \mathcal{D}_{\nu}\mathcal{D}_{\alpha}\pi\mathcal{D}_{\mu}\pi^* \mathcal{D}^{\alpha}\pi^* + \rm{c.c.}\,\,)\, ]\nonumber\\&{} \frac{\omega}{4 \lambda_{\pi}^6}\int d^4x\,\,\sqrt{-g}\,\, \left[\mathcal{D}_{\alpha}\pi\mathcal{D}^{\alpha}\pi \mathcal{D}_{\mu}\pi^*\mathcal{D}_{\nu}\pi^* \left( R^{\mu\nu} - \frac{1}{4}g^{\mu\nu}R \right) + \rm{c.c.}\right]\nonumber\\&{}   -\frac{\omega}{4 \lambda_{\pi}^6}\int d^4x\,\sqrt{-g}\, \mathcal{D}_{\alpha}\pi\mathcal{D}^{\alpha}\pi^* \mathcal{D}_{\mu}\pi \mathcal{D}_{\nu}\pi^*g^{\mu\nu}\,R -\frac{1}{4}\int d^4x F_{\mu\nu}F^{\mu\nu} + S_{\rm{VEG}}
\end{align}
where the last term $S_{\rm{VEG}}$ is the vector Galileon action obtained in Ref.~\cite{2017-Nandi.Shankaranarayanan-JCAP}. 
We have listed the terms in Appendix \eqref{app:D}. We will use this 
action to study the effects of the Galileon term in the early Universe.
\subsection{Fixing the value of $\omega$}
\label{sec:coeffiFixing}

$\lambda_{\pi}$ is the new coupling constant of the model and can only be fixed with observations. As mentioned in Sec. \eqref{sec:CSG}
$\omega$ can take either $+1$ or $-1$. In this subsection, we fix the value of $\omega$ by evaluating the energy density corresponding to the minimal and non-minimal terms in \eqref{action_full_curved} in the Coulomb gauge ($A^0 = 0, \partial_i A^{i} = 0$). 

To make the calculations transparent, we evaluate the energy density 
in the FRW background which includes arbitrary Lapse function $N(t)$: 
\begin{equation}
\label{eq:FRW}
ds^2 = N^2(t) \, dt^2 - a^2(t)(dx^2 + dy^2 + dz^2)
\end{equation}
where $a(t)$ is the scale factor. To satisfy the homogeneity and isotropy of the FRW background, the Galileon scalar electrodynamics must satisfy the condition $A^i = 0$. The equation of motion of $\pi$ corresponding to $S_4^{\min} + S_4^{nm}$ is given by (using Eq.(\ref{eq:S4eomCurved})):
\begin{align}\label{EOM_FRW}
2\dot{\pi}^*\,\ddot{\pi}^*\dot{\pi}\left(\frac{\dot{a}}{a}\right)^2-2(\dot{\pi}^*)^2\,\dot{\pi}\left(\frac{\dot{a}}{a}\right)^2\frac{\dot{N}}{N} +33(\dot{\pi}^*)^2\dot{\pi}\left(\frac{\dot{a}}{a}\right)^3 + (\dot{\pi}^*)^2\ddot{\pi}\left(\frac{\dot{a}}{a}\right)^2\nonumber\\ -(\dot{\pi}^*)^2\dot{\pi}\left(\frac{\dot{a}}{a}\right)^2 \frac{\dot{N}}{N}+ 8(\dot{\pi}^*)^2\dot{\pi}\left(\frac{\ddot{a}}{a}\right)\left(\frac{\dot{a}}{a}\right) -8(\dot{\pi}^*)^2\dot{\pi}\left(\frac{\dot{a}}{a}\right)^2\left(\frac{\dot{N}}{N}\right)= 0
\end{align}

where $H(t) = \dot{a}(t)/a(t)$. Varying the action $ {S}_4^{\rm{min}} + {S}^{\rm{nm}}_4$ with respect to the $g^{00} = N^{-2}(t)$ leads to:
\begin{align}
\delta {S}_4^{\rm{min}} + \delta {S}^{\rm{nm}}_4 &= - \frac{3\omega}{2 \lambda_{\pi}^6} \int d^4 x \frac{a^3}{N^2}
\left[ 79\, ({\dot{\pi}}^*)^2 \,\dot{\pi}\ddot{\pi} \, H(t) 
 + 79\, ({\dot{\pi}})^2 \,\dot{\pi}^*\ddot{\pi}^* \, H(t) \right.\nonumber\\&{}\hspace{1.5cm}\left.
 + 20 \, ({\dot{\pi}}^*)^2 \,(\dot{\pi})^2 H^2(t) 
 +23\, ({\dot{\pi}}^*)^2 \,(\dot{\pi})^2 \left(\frac{\ddot{a}}{a}\right)\right]\,\, \delta N
\end{align}

Using the definition, 
\begin{equation}
T_{0 0} = \frac{2}{\sqrt{-g}}\frac{\delta S}{\delta g^{0 0}} 
= - \frac{N^2}{a^3} \frac{\delta S}{\delta N} \, ,
\end{equation}
we get,
\begin{align}\label{energyd}
\rho =\,\,T^0_0 =  \rho = \frac{3 \omega }{2\,\lambda_{\pi}^6}\,\, \left[ \frac{79}{2}\,\partial_{0}N_1 \,H + 20\,N_1 \,H^2 + 23\,\frac{\ddot{a}}{a}\,N_1\,\,\right]
\end{align}
where $N_1 = |\dot{\pi}|^4$.  In order to get the value of $\omega$ we simplify the analysis further by setting $N(t) = 1$ and taking the limit, $\pi = \pi^*$. Thus, the energy density and equation of motion reduce to
\begin{align}
\label{energy_back}
& \rho =  \frac{3 \omega}{2 \lambda_{\pi}^6} \left[ 2\times79\, ({\dot{\pi}})^3 \,\ddot{\pi} \,H + 20 \, (\dot{\pi})^4 \,H^2+23\, (\dot{\pi})^4 \,(H^2 + \dot{H})\right] \\
\label{eom_back}
& \ddot{\pi} +\frac{41}{3}\,\dot{\pi}H + \frac{8}{3}\,\dot{\pi}\left(\frac{\dot{H}}{H}\right) = 0
\end{align}
Substituting Eq.~(\ref{eom_back}) in Eq.~(\ref{energy_back}), we get
\begin{align}
\rho = -\frac{\omega \dot{\pi}^4\,H^2}{2 \lambda_{\pi}^6}\left[ 6349 + 1195\,\frac{\dot{H}}{H^2} \right]
\end{align}
Note that $\dot{H}/H^2 = - \epsilon$ which is a slow-roll parameter. 
During inflation, when the non-linear terms contribution can not be ignored, $\epsilon < 1$, hence the quantity in the square bracket, is positive. The condition that the energy density is positive implies that $\omega = -1$. 

%the sign of coefficient $\lambda^6$ must be negative because energy density is a positive definite quantity. Hence the original coefficient $A_2$ should be negative. So we can set $\omega = -1$, the expression for energy density becomes
%\begin{align}\label{rho_final}
%\rho = \frac{ \dot{\pi}^4\,H^2}{2 \lambda_{\pi}^6}\left[ 6349 + 1195\,\frac{\dot{H}}{H^2} \right]
%\end{align}
%After fixing the sign of the coefficient at the background. We write the energy density expression in terms of $\pi$ and $\pi^*$ and some convenient quantities which will be useful for further calculations. These quantities are defined as
%\begin{align}
%\alpha_1 = (\dot{\pi}^*)^2 \dot{\pi}\,\ddot{\pi}\,a, \,\,\, \alpha_2 = (\dot{\pi})^2 \dot{\pi}^*\,\ddot{\pi}^*\,a,\,\,\, \alpha_3 = (\dot{\pi}^*)^2 \,(\dot{\pi})^2\,\dot{a}, \,\,\, \alpha_4 = (\dot{\pi}^*)^2 \,(\dot{\pi})^2\,a
%\end{align}
%using these relations we get, $ (\dot{\pi}^*)^2 \,(\dot{\pi})^2\,\ddot{a}\,a = a\,\partial_{0}\alpha_3 - 2\dot{a}\,(\alpha_1 + \alpha_2)$ and $\partial_{0}\alpha_4 = 2\,\alpha_1 + 2\,\alpha_2 + \alpha_3$, so the energy density in Eq. (\ref{energyd}) becomes
%\begin{align}\label{rho}
%\rho = \frac{3 \omega}{2 \lambda_{\pi}^6 \, a} \left[ 33\, \alpha_1\,H + 33\, \alpha_2\,H + 20\,\alpha_3 \,H+ 23\,\partial_{0}\alpha_3\right]
%\end{align}

%%%%%%55
\section{Applications to Early-Universe}
\label{sec:energy_pressure}

\begin{figure}[h]
	\centering
	\includegraphics[width=0.85\textwidth]{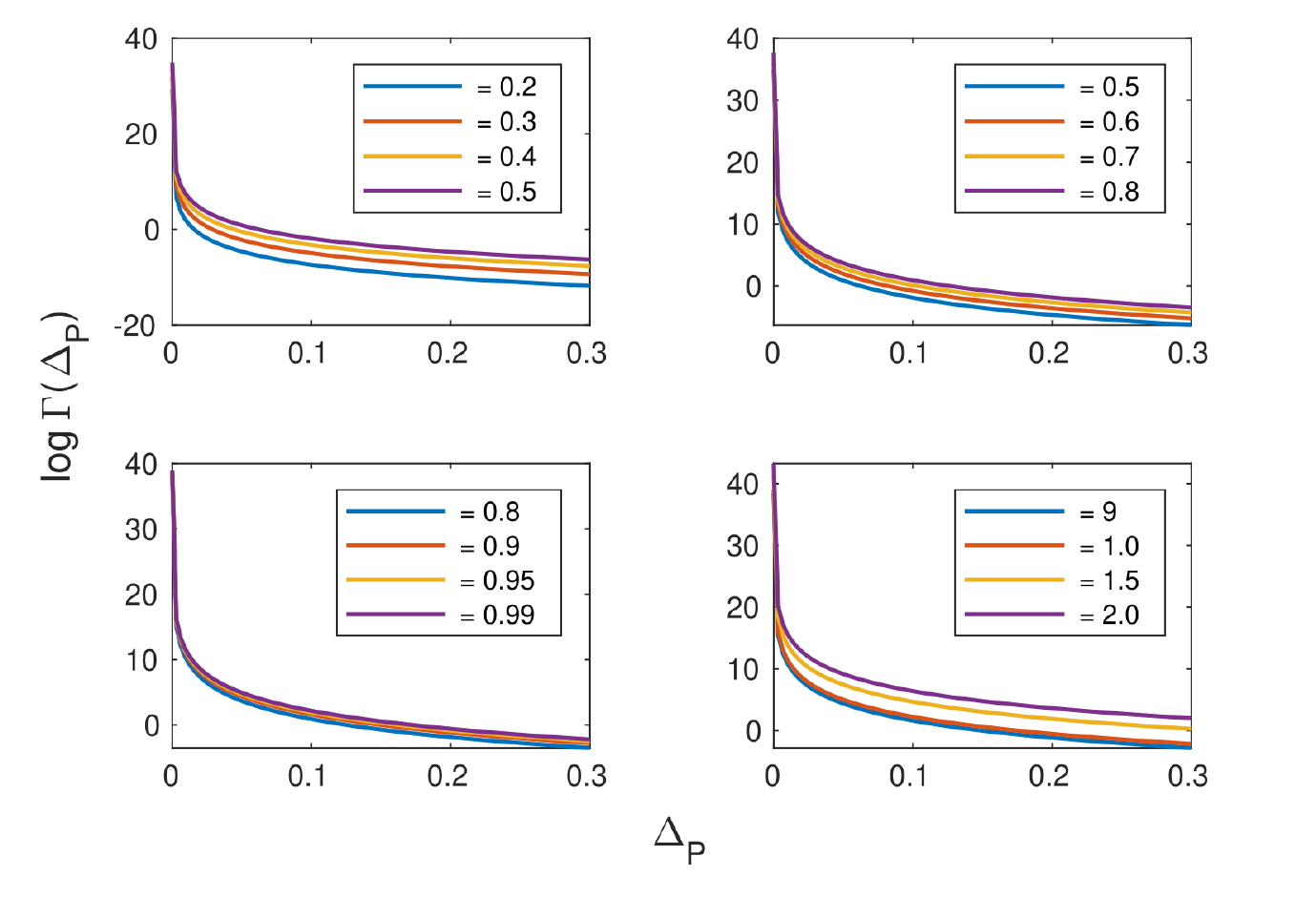}
	\caption{Plot of $\log\Gamma(\Delta_P) $ versus $\Delta_P$ for different values of $\delta_P$. Note that for a range of values of the two parameters, $\Gamma$ is also zero.}
	\label{fig:Plot}
\end{figure}

To extract interesting features of the Galileon model, in this section, we show that the action \eqref{action_full_curved} leads to accelerated expansion in the early Universe. We obtained the energy density of the Galileon field in \eqref{energy_back}. Using the same procedure, the 
pressure of the Galileon field is given by:
\begin{align}\label{pressure}
p &= \frac{\omega }{2\,\lambda_{\pi}^6}\,\,[\,\,17 \dot{N_1} \,H- 72\, N_1 \,H^2+ 8\, \frac{\ddot{a}}{a}\,N_1  + 3\,\ddot{N_1} \,\,]
%p  = - \frac{1}{\lambda_{\pi}^6}\,\,[\,\, 
%6 N_2 \,H + 6 N_3 \,H -3 6 N_1 \,H + 
%3\, \partial_t \left(N_2 \, a\right) + 
%3\, \partial_{t} \left(N_3 \, a \right) + 
%4\, \partial_{t} \left(N_1 \dot{a} \right)\,\,]
\end{align}
%%
%where 
%\begin{align}\label{N123}
%N_2 &= (\dot{\pi}^*)^2 \dot{\pi}\,\ddot{\pi};~~
%N_3 =(\dot{\pi})^2 \dot{\pi}^*\,\ddot{\pi}^*;
%\end{align}
%%
Taking the case that the phase variation of the complex field is almost constant, the Friedmann equation becomes:
\begin{align}\label{eq:slow-roll}
1 = \frac{\dot{\pi}^4}{6\,M_P^2\,\lambda_{\pi}^6}\left[ 6349 -1195\,\epsilon \right]
\end{align}
Inverting the above equation, we get 
\begin{align}\label{eq:epsilon}
\epsilon = \frac{6349}{1195}\left[1 - \Gamma\right] \quad 
\mbox{where} \quad \Gamma = 
\frac{6\,M_P^2\,\lambda_{\pi}^6}{6349\,\dot{\pi}^4}\,
\end{align}
In order for $\epsilon$ to be less than unity,  $\Gamma \simeq 1$. 
To identify the parameter space of $(\dot{\pi}, \lambda_{\pi})$ that can lead to $\Gamma \simeq 1$, we define $\dot{\pi} \equiv \Delta_P \,M_P^2, \lambda_{\pi} \equiv \delta_P \, M_P$ where $\Delta_P$ and $\delta_P$ are dimensionless quantities. \eqref{fig:Plot} contains 
the plot of $\log \Gamma$ as a function of $\Delta_P$ for different values of $\delta_P$. We notice the following: First, the value of $\Gamma$ is weakly dependent on $\delta_P$. In other words, for different values of $\delta_P$, the value of $\Gamma$ is almost the same. Second, for a large values of $\Delta_P$, $\Gamma$ is almost close to unity and hence, the slow-roll parameter is very small. More specifically, for $\Delta_P$ in the range $10^{-5}$ to $0.3$, the slow-roll parameter is of the order of $0.1$. Thus, for large parameter range, the model can lead to extended period of inflation. 

\section{Conclusions and Discussions}
\label{sec:conc}

In this work, we constructed Galileon scalar electrodynamics action, which preserves Galilean symmetry and local gauge invariance. Due to the complex scalar field, the number of Galilean symmetry 
invariant terms are reduced. For an earlier result, in a different context, see Ref. \cite{Goon:2012mu}. In the flat space-time, we have explicitly shown that the equations of motion are second order. 
In curved space-time, due to the non-commutative nature of the covariant derivatives, the minimal coupling of the matter and gravity term leads to higher-derivatives in the equation of motion. We introduced non-minimal coupling terms to the Galileon field that makes the equations of motion second-order in an arbitrary curved space-time. The non-minimal coupling term is different compared that the one used in the literature for the real scalar field~\cite{Deffayet2009}. However, in the real scalar field limit, the complex scalar Galileon action is identical to the real scalar Galileon action obtained by Deffayet et al~\cite{Deffayet2009}.

As an application of the model, we considered the case when the Galileon scalar electrodynamics dominated the early Universe. We have shown that for a range of parameters, the model leads to inflation. One possible application of the model is to study the perturbations generated during the inflation dominated by the model. 
Since the model breaks conformal invariance, this can generate primordial magnetic fields~\cite{2017-Nandi.Shankaranarayanan-JCAP}. We hope to report this in more detail shortly.

The model has two free parameters -- $\lambda_{\pi}$ and $\lambda_{VG}$. In the flat FRW background, the electromagnetic field vanishes and hence the scale of inflation fixes $\lambda_{\pi}$. 
The other parameter $\lambda_{VG}$ will play a critical role in the first order perturbations. In principle, the two parameters can be different.

The model presented here can be part of the scalar-vector-galileon (SVG) theories. However, as mentioned earlier, not all SVG gravity theories are gauge-invariant, while our model is gauge-invariant~\cite{2018-Heisenberg.etal-PRD}. 
The complexification of a gauge-invariant subset of the SVG theories should be identical to our model with an appropriate choice of coefficients $A_{i}, B_{i}, C_{i}, D_{i}$ (for $i=1,2$) and $D_{3}$. We hope to report this elsewhere.

\begin{acknowledgments}
The authors wish to thank Debottam Nandi for useful discussions. The authors thank the anonymous referee for raising a point which clarified an important issue in the work. The MHRD fellowship at IIT Bombay financially supports AK. This work is supported by the ISRO-Respond grant. Further, we thank Kasper Peeters for his useful program Cadabra~\cite{2007-Peeters1-arXiv,2007-Peeters2-arXiv}, and useful algebraic calculations with it.
\end{acknowledgments}

\appendix
\section{Fixing the coefficients of $\mathcal{L}_4$}\label{Generic}

In this Appendix, we fix all the coefficients in the Lagrangian \eqref{L4}. Demanding that the Lagrangian is invariant under the Galileon transformation \eqref{eq:GalileanShift} leads to the 
following constraint:
\begin{align}\label{LCGGT}
2A_1 \,\partial_{\mu} \partial^{\mu}\pi^* \partial_{\nu} \partial^{\nu}\pi^* \partial_{\alpha}\pi a^{\alpha} 
+ A_2 \,\partial_{\mu}\partial^{\mu}\pi^* \partial_{\nu} \partial^{\nu}\pi \partial_{\alpha}\pi^* a^{\alpha} 
+  B_1\,(\partial_{\mu}\partial^{\mu}\pi^* \partial^{\nu} \partial^{\alpha}\pi \partial_{\nu}\pi^* a_{\alpha} + \partial_{\mu}\partial^{\mu}\pi^* \partial^{\nu} \partial^{\alpha}\pi \partial_{\alpha}\pi {a_{\nu}}^*) \nonumber\\
+ 2B_2\,\partial_{\mu}\partial^{\mu}\pi^* \partial^{\nu} \partial^{\alpha}\pi^* \partial_{\nu}\pi a_{\alpha} 
+ 2 C_1\,\partial_{\mu}\partial_{\nu} \pi^*\partial^{\mu}\partial^{\nu}\pi^* \partial^{\alpha}\pi a_{\alpha} 
+ C_2\, \partial_{\mu}\partial_{\nu}\pi^* \partial^{\mu}\partial^{\nu}\pi \partial^{\alpha}\pi^* a_{\alpha} \nonumber\\
+ 2 D_1\, \partial^{\mu}\partial^{\nu}\pi^* \partial_{\nu}\partial_{\alpha}\pi^* \partial_{\mu}\pi a^{\alpha}
+ D_2 \,\partial^{\mu}\partial^{\nu}\pi^* \partial_{\nu}\partial_{\alpha}\pi \partial_{\mu}\pi^* a^{\alpha}
+ D_3\,\partial^{\mu}\partial^{\nu}\pi \partial_{\nu}\partial_{\alpha}\pi^* \partial_{\mu}\pi^* a^{\alpha} +c.c. = 0 \, .
\end{align}    
Here, c.c. denotes the complex conjugate, we have considered only first order terms in $a^{\mu}$ and have ignored the quadratic terms of the infinitesimal shift parameters. All the first order terms in Eq.~(\ref{LCGGT}) either do not contribute or contribute total derivative terms. After integrating by parts (and ignoring the total derivative terms as they do not contribute to the action) and collecting similar terms, we get
\begin{align}\label{a1}
\left[(2A_1 - B_1)\,(\Box \pi^*)^2\partial_{\alpha}\pi + (A_2 -2 B_2 - B_1)\,\Box \pi^*\Box \pi \partial_{\alpha}\pi^*
- ( B_1 +2C_1)\,\partial^{\nu}\Box\pi^*\partial_{\nu}\pi^* \partial_{\alpha}\pi\right. \nonumber\\
\left. - (B_1 + D_3)\,\partial^{\mu}\Box \pi \partial_{\mu}\pi^*\partial_{\alpha}\pi^*+ 
(C_2 -2 D_1 -D_3)\, \partial_{\mu}\partial_{\nu}\pi^* \partial^{\mu}\partial^{\nu}\pi \partial_{\alpha}\pi^* \right. \nonumber\\
\left.
- (2D_1 + 2B_2)\,\partial^{\mu}\Box \pi^*\partial_{\mu}\pi \partial_{\alpha}\pi^* - (2C_1 - D_2)\,\partial_{\mu}\partial_{\nu}\pi^* \partial^{\nu}\partial_{\alpha}\pi \partial^{\mu}\pi^*\right]a^{\alpha} = 0
\end{align}
Only way to satisfy the above equation for an arbitrary $\pi$ and $\pi^*$ is that all the coefficients vanish. This leads to the following relations:
\begin{align}\label{Coeff}
B_1 &= 2 A_1;~C_1 = - A_1 \nonumber\\
B_2 &= -A_1 + \frac{A_2}{2} = -E\nonumber\\
C_2 &= - A_2;~D_2 = -2 A_1 \nonumber\\
D_1 &= A_1 - \frac{A_2}{2} = E\nonumber\\
D_3 &= -2 A_1 \, .
\end{align}
It is straightforward to see from the above expressions that all 
the parameters can be written in terms of two  parameters $A_1$ and $A_2$. Substituting the coefficients (\ref{Coeff}) in Eq.~(\ref{L4}) leads to the following generic complex scalar Lagrangian $\mathcal{L}_{4}$;
 \begin{align}\label{LCGG}
{\mathcal{L}}_{4}&{} = A_1 [\,\, (\,\,(\Box \pi)^2 \partial_{\alpha}\pi^* \partial^{\alpha}\pi^* +  2 \Box \pi\partial_{\nu}\pi\partial^{\nu} \partial^{\alpha}\pi^* \partial_{\alpha}\pi^* - \partial_{\mu}\partial_{\nu}\pi \partial^{\mu}\partial^{\nu}\pi \partial^{\alpha}\pi^* \partial_{\alpha}\pi^* + \rm{c.c.}\,\,) \nonumber \\&{}-2 \partial_{\mu}\pi^*\partial^{\mu}\partial^{\nu}\pi \partial_{\nu}\partial_{\alpha}\pi^* \partial^{\alpha}\pi ]
 + A_2 [\Box \pi^* \Box \pi \partial_{\alpha}\pi^* \partial^{\alpha}\pi  - \partial_{\mu}\partial_{\nu}\pi^*\partial^{\mu}\partial^{\nu}\pi \partial^{\alpha}\pi^* \partial_{\alpha}\pi ] \nonumber \\&{}
+ E [\partial_{\mu}\pi^*\partial^{\mu}\partial^{\nu}\pi \partial_{\nu}\partial_{\alpha}\pi \partial^{\alpha}\pi^* -\Box \pi\partial_{\nu}\pi^* \partial^{\nu} \partial^{\alpha}\pi \partial_{\alpha}\pi^* + \rm{c.c.}  ]
\end{align}    
Now, there are only two free parameters $A_1$ and $A_2$ in the Lagrangain.     
The action for the above generic Lagrangian (\ref{LCGG}) is given by,
\begin{equation}\label{S_GCC}
{S}_{4} = \int d^4 x~\mathcal{L}_{4}
\end{equation}

We can identify the generic action (\ref{S_GCC}) as the summation of the following actions:
\begin{equation}
\begin{split}\label{S_i}
S^{(1)}_{4} &= A_1 \int d^4 x \; (\, \partial_{\mu} \partial^{\mu}\pi \partial_{\nu} \partial^{\nu}\pi \partial_{\alpha}\pi^* \partial^{\alpha}\pi^* + \rm{c.c.} \,\,)\\
{S}^{(2)}_{4} &= A_2 \int d^4 x \; \partial_{\mu} \partial^{\mu}\pi^* \partial_{\nu} \partial^{\nu}\pi \partial_{\alpha}\pi^* \partial^{\alpha}\pi\\
{S}^{(3)}_{4} &= 2 A_1 \int d^4 x \; (\, \partial_{\mu}\partial^{\mu}\pi\partial_{\nu}\pi \partial^{\nu} \partial^{\alpha}\pi^* \partial_{\alpha}\pi^* + \rm{c.c.}\,\,)\\
{S}^{(4)}_{4} &= -E \int d^4 x \; (\,\partial_{\mu}\partial^{\mu}\pi\partial_{\nu}\pi^* \partial^{\nu} \partial^{\alpha}\pi \partial_{\alpha}\pi^* + \rm{c.c.} \,\,)\\
{S}^{(5)}_{4} &= -A_1 \int d^4 x \; (\, \partial_{\mu}\partial_{\nu}\pi \partial^{\mu}\partial^{\nu}\pi \partial^{\alpha}\pi^* \partial_{\alpha}\pi^* + \rm{c.c.}\,\,)\\
{S}^{(6)}_{4} &= -A_2 \int d^4 x \; \partial_{\mu}\partial_{\nu}\pi^*\partial^{\mu}\partial^{\nu}\pi \partial^{\alpha}\pi^* \partial_{\alpha}\pi\\
{S}^{(7)}_{4} &= E \int d^4 x \; (\, \partial_{\mu}\pi\partial^{\mu}\partial^{\nu}\pi^* \partial_{\nu}\partial_{\alpha}\pi^* \partial^{\alpha}\pi + \rm{c.c.}\,\,)\\
{S}^{(8)}_{4} &=  -2 A_1 \int d^4 x \; \partial_{\mu}\pi^*\partial^{\mu}\partial^{\nu}\pi^* \partial_{\nu}\partial_{\alpha}\pi \partial^{\alpha}\pi\\
{S}^{(9)}_{4} &= -2 A_1\int d^4 x \; \partial_{\mu}\pi^*\partial^{\mu}\partial^{\nu}\pi \partial_{\nu}\partial_{\alpha}\pi^* \partial^{\alpha}\pi
\end{split}
\end{equation}
where
\begin{equation}\label{S4}
{S}_{4} = \sum_{i=1}^{9} \; {S}^{(i)}_{4}
\end{equation}
Having obtained the generic 4th order Galileon action  (\ref{S_i}), our next step is to calculate the equations of motion of $\pi$ ( for $\pi^*$ is straight forward). We define the quantity $\mathcal{E}_i$ as the variation of the ${S^{(i)}}_{CGG}$ with respect to $\pi$ (with respect to $\pi^*$ for ${\mathcal{E}_i}^*$) as, 
\begin{equation}\label{E_i}
\mathcal{E}^i_4 = \frac{\delta {S}^{(i)}_{4}}{\delta \pi}
\end{equation}
Hence, equation of motion for the generic action (\ref{S4}) is given by,
\begin{align}\label{E_CGG}
\mathcal{E}_{4} = \sum_{i=1}^{9} \; \mathcal{E}_4^i
\end{align}
Using Eq.~(\ref{E_i}), the equations of motion corresponding to nine actions are
\begin{align}
\label{eq:e1}
\mathcal{E}^{(1)}_4 &= 2A_1\,[\partial_{\mu}\partial^{\mu}\Box \pi\, \partial_{\alpha}\pi^*\partial^{\alpha}\pi^* + 4\,\partial_{\mu}\Box \pi\, \partial_{\alpha}\pi^*\partial^{\mu}\partial^{\alpha}\pi^*  + 2\, \Box \pi\, \partial_{\alpha} \Box \pi^*\,\partial^{\alpha}\pi^*+ 2 \,\Box \pi\, \partial_{\mu}\partial_{\alpha}\pi^*\,\partial^{\mu}\partial^{\alpha}\pi^* \nonumber\\ &\hspace{1cm}- 2\,\partial^{\alpha}\Box \pi^*\, \Box \pi^*\, \partial_{\alpha}\pi - (\Box \pi^*)^2 \,\Box \pi]
\\
\vspace{1cm}\nonumber\\
\label{eq:e2}
\mathcal{E}^{(2)}_4 &= A_2\,[\partial_{\mu}\partial^{\mu}\Box \pi^*\, \partial_{\alpha}\pi\partial^{\alpha}\pi^* + 2\,\partial_{\mu}\Box \pi^*\, \partial^{\mu}\partial^{\alpha}\pi \partial_{\alpha}\pi^*  + 2\, \partial_{\nu}\Box \pi^*\, \partial_{\alpha}\pi \partial^{\nu}\partial^{\alpha}\pi^* +  \Box \pi^*\, \partial_{\alpha} \Box \pi \,\partial^{\alpha}\pi^* \nonumber \\&{}
+ \Box \pi^*\,\partial^{\alpha}\Box \pi^*\, \partial_{\alpha} \pi + 2\, \Box \pi^*\, \partial_{\nu}\partial_{\alpha}\pi \partial^{\nu}\partial^{\alpha}\pi^* - \partial_{\alpha}\Box \pi^*\,\Box \pi \,\partial^{\alpha}\pi^* - \Box\pi^*\, \partial_{\alpha} \Box \pi\, \partial^{\alpha}\pi^* - (\Box \pi^*)^2\, \Box \pi]
\\
\vspace{1cm}\nonumber\\
\label{eq:e3}
\mathcal{E}^{(3)}_4 &= 2A_1\,[\,2\,\partial^{\alpha}\partial^{\nu}\Box \pi^*\, \partial_{\nu}\pi^*\,\partial_{\alpha}\pi + 2\,\partial^{\nu}\Box \pi^*\, \partial^{\alpha}\pi\,\partial_{\nu}\partial_{\alpha}\pi^* + 2\partial^{\alpha}\Box \pi^* \Box \pi^*\,\partial_{\alpha}\pi + (\Box \pi^*)^2\, \Box \pi \nonumber\\&{}+ 2\,\partial^{\mu}\partial^{\nu}\partial^{\alpha}\pi^*\,\partial_{\mu}\partial_{\nu}\pi\, \partial_{\alpha}\pi^*
+ 2\, \partial^{\mu}\partial^{\nu}\partial^{\alpha}\pi^*\, \partial_{\mu}\partial_{\alpha}\pi^*\, \partial_{\nu}\pi +  2\,\partial^{\nu}\partial^{\alpha}\pi^*\,\partial_{\mu}\partial_{\nu}\pi\, \partial^{\mu}\partial_{\alpha}\pi^* - \Box \pi \,\partial^{\nu}\partial^{\alpha}\pi^*\, \partial_{\nu}\partial_{\alpha}\pi^*]
\\
\vspace{1cm}\nonumber\\
\label{eq:e4}
\mathcal{E}^{(4)}_4 &= -E\,[\,2\,\partial^{\nu}\Box \pi^*\, \partial_{\nu}\partial_{\alpha}\pi\,\partial^{\alpha} \pi^* + 4\,\partial^{\mu}\partial^{\nu}\partial^{\alpha}\pi\, \partial_{\mu}\partial_{\nu}\pi^*\,\partial_{\alpha}\pi^* +   2\,\partial^{\mu}\partial^{\nu}\pi^*\,\partial_{\nu}\partial_{\alpha}\pi\, \partial_{\mu}\partial^{\alpha}\pi^* 
+ 2\,\partial_{\alpha}\partial_{\nu}\Box \pi \,\partial^{\nu}\pi^*\, \partial^{\alpha}\pi^*\, \nonumber\\&{}+ 2\,\partial^{\nu}\Box \pi\, \partial_{\alpha}\partial_{\nu}\pi^*\, \partial^{\alpha}\pi^* + 2\,\partial^{\nu}\Box \pi\, \Box \pi^*\, \partial_{\nu}\pi^* +
2\,\partial^{\alpha}\Box \pi^*\,\Box \pi \,\partial_{\alpha}\pi^*  + \Box \pi\,\partial_{\nu}\partial_{\alpha}\pi^* \,\partial^{\nu}\partial^{\alpha}\pi^* + (\Box \pi^*)^2\,\Box \pi \nonumber\\&{}- \partial_{\nu}\Box \pi^*\, \partial^{\nu}\partial^{\alpha}\pi^*\, \partial_{\alpha}\pi - 2\,\Box \pi^*\, \partial^{\alpha}\Box \pi^* \,\partial_{\alpha}\pi  - 2 \,\Box \pi^*\,\partial^{\nu}\partial^{\alpha}\pi^*\,\partial_{\nu}\partial_{\alpha}\pi - \partial_{\alpha} \Box \pi^* \,\partial^{\nu}\partial^{\alpha}\pi^*\,\partial_{\nu}\pi]
\\
\vspace{1cm}\nonumber\\
\label{eq:e5}
\mathcal{E}^{(5)}_4 &= -2A_1\,[\,\partial_{\nu}\partial^{\nu} \Box\pi\, \partial_{\alpha}\pi^*\,\partial^{\alpha}\pi^* + 2\,\partial^{\nu}\Box \pi\, \partial^{\alpha}\pi^*\,\partial_{\nu}\partial_{\alpha}\pi^* + 2\,\partial^{\mu}\Box \pi\, \partial^{\alpha}\pi^*\,\partial_{\mu}\partial_{\alpha}\pi^* + 2\,\partial^{\mu}\partial^{\nu} \pi\, \partial_{\nu}\partial_{\alpha}\pi^*\,\partial_{\mu}\partial^{\alpha}\pi^* \nonumber\\&{}+ 2\,\partial^{\mu}\partial^{\nu} \pi\, \partial^{\alpha}\pi^*\,\partial_{\nu}\partial_{\mu}\partial_{\alpha}\pi^*\,  -2\,\partial_{\nu}\partial_{\mu}\partial_{\alpha}\pi^*\, \partial^{\mu}\partial^{\nu} \pi^*\, \partial^{\alpha}\pi - \Box \pi\,\partial_{\mu}\partial_{\nu}\pi^*\, \partial^{\mu}\partial^{\nu}\pi^* ]
\\
\vspace{1cm}\nonumber\\
\label{eq:e6}
\mathcal{E}^{(6)}_4 &= -A_2\,[\,\partial_{\nu}\partial^{\nu}\Box \pi^*\, \partial_{\alpha}\pi\,\partial^{\alpha}\pi^* + \partial_{\nu}\Box \pi^*\, \partial^{\nu}\partial^{\alpha}\pi\, \partial_{\alpha}\pi^* + \partial_{\nu}\Box \pi^*\, \partial_{\alpha} \pi\,\partial^{\nu}\partial^{\alpha}\pi^* + \partial_{\mu}\Box \pi^*\, \partial^{\mu}\partial^{\alpha}\pi \,\partial_{\alpha}\pi^* \nonumber\\&{}+ \partial_{\mu}\partial_{\nu}\pi^*\, \partial^{\mu}\partial^{\alpha}\pi\,\partial^{\nu} \partial_{\alpha}\pi^* + \partial_{\mu}\Box \pi^* \,\partial_{\alpha}\pi\, \partial^{\mu}\partial^{\alpha}\pi^* 
+ \partial_{\mu}\partial_{\nu}\pi^* \,\partial^{\nu}\partial^{\alpha}\pi\, \partial^{\mu}\partial_{\alpha}\pi^* +  \partial_{\mu}\partial_{\nu}\pi^*\,\partial_{\alpha}\pi\, \partial^{\mu}\partial^{\nu}\partial^{\alpha}\pi^* \nonumber\\&{}- \partial^{\mu}\partial^{\nu}\pi \,\partial^{\alpha}\pi^*\, \partial_{\alpha}\partial_{\mu}\partial_{\nu}\pi^*- \Box \pi^*\, \partial_{\mu}\partial_{\nu}\pi^* \,\partial^{\mu}\partial^{\nu}\pi ]
%\\
%\vspace{1cm}\nonumber
\end{align}
\begin{align}
\label{eq:e7}
\mathcal{E}^{(7)}_4 &= E\,[\,2\,\partial^{\mu}\partial^{\alpha}\Box \pi\, \partial_{\mu}\pi^*\,\partial_{\alpha}\pi^* + 4\,\partial^{\mu}\partial^{\nu}\partial^{\alpha}\pi\, \partial_{\nu}\partial_{\mu}\pi^*\,\partial_{\alpha}\pi^* + 2\,\partial_{\mu}\Box \pi\, \Box \pi^*\,\partial^{\mu}\pi^* + 2\,\partial_{\alpha}\Box \pi\, \partial_{\mu}\pi^*\, \partial^{\mu}\partial^{\alpha}\pi^* \nonumber\\&{} + 2\,\Box\pi^* \,\partial_{\nu}\partial_{\alpha}\pi\, \partial^{\nu}\partial^{\alpha}\pi^* + 2\,\partial^{\nu}\Box\pi^*\, \partial_{\nu}\partial_{\alpha}\pi\, \partial^{\alpha}\pi^*  + 2\,\partial_{\nu}\partial_{\alpha} \pi \,\partial^{\nu}\partial^{\mu} \pi^*\, \partial_{\mu}\partial^{\alpha} \pi^* + 2\,\partial_{\nu}\partial_{\alpha} \pi\,\partial_{\mu} \pi^* \,\partial^{\nu}\partial^{\mu}\partial^{\alpha} \pi^* \nonumber\\&{} - 2\,\partial^{\nu}\Box \pi^* \,\partial_{\nu}\partial_{\alpha} \pi^*\, \partial^{\alpha} \pi -2\,\partial^{\mu}\partial^{\nu} \pi^*\, \partial_{\mu}\partial_{\nu}\partial_{\alpha} \pi^*\, \partial^{\alpha} \pi - 2\,\partial^{\mu}\partial^{\nu} \pi^* \,\partial_{\nu}\partial_{\alpha} \pi^*\, \partial_{\mu}\partial^{\alpha} \pi] \\
%%%%
\vspace{1cm}\nonumber\\
\label{eq:e8}
\mathcal{E}^{(8)}_4 &= -2\,A_1\,[\,\partial^{\mu}\partial^{\alpha}\Box \pi^*\, \partial_{\mu}\pi^*\,\partial_{\alpha}\pi + \partial^{\mu}\Box\pi^*\, \partial_{\alpha}\partial_{\mu}\pi^*\,\partial^{\alpha}\pi + \partial^{\mu}\Box \pi^* \,\Box \pi\,\partial_{\mu}\pi^* + 2 \,\partial_{\alpha}\partial_{\mu}\partial_{\nu}\pi^*\, 
\partial^{\nu}\partial^{\mu}\pi^*\, \partial^{\alpha}\pi \nonumber \\&{}\hspace{1cm}+ \partial_{\mu}\partial_{\nu}\pi^* \,
\partial^{\nu}\partial^{\mu}\pi^*\, \Box\pi
+ \partial^{\alpha}\partial^{\mu}\partial^{\nu}\pi^* \,\partial_{\mu}\pi^*\, \partial_{\nu}\partial_{\alpha}\pi + \partial^{\mu}\partial^{\nu}\pi^*\, \partial_{\mu}\partial_{\alpha}\pi^* \,\partial_{\nu}\partial^{\alpha}\pi + \partial^{\mu}\partial^{\nu}\pi^*\,\partial_{\mu}\pi^*\,\partial_{\nu}\Box \pi  \nonumber\\&{}\hspace{1cm} - \partial^{\mu}\partial^{\nu}\partial^{\alpha}\pi^*\, \partial_{\nu}\partial_{\alpha} \pi\, \partial_{\mu}\pi^*
-\partial^{\mu}\partial^{\nu} \pi^* \partial_{\nu}\Box \pi \partial_{\mu}\pi^* - \partial^{\mu}\partial^{\nu}\pi^* \partial_{\nu}\partial_{\alpha} \pi \partial^{\alpha}\partial_{\mu} \pi^*]
\\
\vspace{1cm}\nonumber\\
\label{eq:e9}
\mathcal{E}^{(9)}_4 &= -2\,A_1\,[\,\partial^{\mu}\partial^{\alpha}\Box \pi^*\, \partial_{\mu}\pi^*\,\partial_{\alpha}\pi + \partial^{\mu}\partial^{\nu}\partial^{\alpha}\pi^*\,\partial_{\nu}\partial_{\mu}\pi^*\, \partial_{\alpha}\pi + \partial^{\mu}\partial^{\nu}\partial^{\alpha} \pi^*\, \partial_{\mu} \pi^*\,\partial^{\nu}\partial^{\alpha}\pi + \partial_{\alpha}\Box\pi^*\,\Box\pi^* \,
\partial^{\alpha}\pi \nonumber\\&{}\hspace{1cm} + \partial_{\alpha}\partial_{\nu}\pi^*\, 
\partial^{\nu}\Box \pi^*\, \partial^{\alpha}\pi+ \partial_{\nu}\partial_{\alpha}\pi^*\, 
\Box \pi^*\, \partial^{\nu}\partial^{\alpha}\pi 
+ \partial_{\alpha}\Box \pi^*\, \partial_{\mu}\pi^*\,\partial^{\mu}\partial^{\alpha}\pi + 
-\partial^{\mu}\partial^{\nu} \pi\, \partial_{\nu}\Box \pi^*\, \partial_{\mu}\pi^* 
]
\end{align}
Thus, the equation of motion for the generic 4th order action \eqref{S4} is:
\begin{align}\label{eq:Ecgg}
\mathcal{E}_{4} &= \left(A_1 + \frac{A_2}{2}\right)[ \,\,\Box \pi\, \partial_{\mu}\partial_{\nu}\pi^*\, \partial^{\mu} \partial^{ \nu}\pi^* + 2\,\Box \pi^*\, \partial_{\mu}\partial_{\nu}\pi \,\partial^{\mu} \partial^{ \nu}\pi^*  -2\,\partial_{\mu}\partial_{\nu}\pi^*\, \partial^{\nu}\partial^{\alpha}\pi\,\partial^{\mu}\partial_{\alpha}\pi^*  -(\,\Box \pi^*)^2\,\Box \pi\,]
\end{align}
To compare with the earlier results~\cite{2009-Nicolis-PRD}, we discuss special cases by taking some specific values of the coefficients $A_1$ and $A_2$.\\
\\
\textbf{Case 1: $A_1 = 0$}
Equations of motion for $\pi$ is given by
\begin{align}\label{eq:case1}
\mathcal{E}_4^{A_1 = 0} &= \frac{A_2}{2}[ \,\,\Box \pi\, \partial_{\mu}\partial_{\nu}\pi^*\, \partial^{\mu} \partial^{ \nu}\pi^* + 2\,\Box \pi^*\, \partial_{\mu}\partial_{\nu}\pi \,\partial^{\mu} \partial^{ \nu}\pi^*  -2\,\partial_{\mu}\partial_{\nu}\pi^*\, \partial^{\nu}\partial^{\alpha}\pi\,\partial^{\mu}\partial_{\alpha}\pi^*  -(\,\Box \pi^*)^2\,\Box \pi\,]
\end{align}
which in the limit $\pi = \pi^*$ gives
\begin{align}\label{eq:case1_scalarlimit}
\mathcal{E}_4^{A_1 = 0} &= \frac{A_2}{2}\,[\, 3\, \Box \pi \,\partial_{\mu}\partial_{\nu}\pi\, \partial^{\mu} \partial^{ \nu}\pi - 2\,\partial_{\mu}\partial_{\nu}\pi\, \partial^{\nu}\partial^{\alpha}\pi\,\partial^{\mu}\partial_{\alpha}\pi -(\Box \pi)^3\,\,]
\end{align}
\textbf{Case 2: $A_2 = 0$}
Equations of motion for $\pi$ is given by
\begin{align}\label{eq:case2}
\mathcal{E}_4^{A_2 = 0} &= A_1[ \,\,\Box \pi\, \partial_{\mu}\partial_{\nu}\pi^*\, \partial^{\mu} \partial^{ \nu}\pi^* + 2\,\Box \pi^*\, \partial_{\mu}\partial_{\nu}\pi \,\partial^{\mu} \partial^{ \nu}\pi^*  -2\,\partial_{\mu}\partial_{\nu}\pi^*\, \partial^{\nu}\partial^{\alpha}\pi\,\partial^{\mu}\partial_{\alpha}\pi^*  -(\,\Box \pi^*)^2\,\Box \pi\,]
\end{align}
which in the limit $\pi = \pi^*$ gives
\begin{align}\label{eq:case2_scalarlimit}
\mathcal{E}_4^{A_2 = 0} &= A_1\,\left[\, 3\, \Box \pi \,\partial_{\mu}\partial_{\nu}\pi\, \partial^{\mu} \partial^{ \nu}\pi - 2\,\partial_{\mu}\partial_{\nu}\pi\, \partial^{\nu}\partial^{\alpha}\pi\,\partial^{\mu}\partial_{\alpha}\pi -(\Box \pi)^3\,\,\right]
\end{align}
\textbf{Case 3: $A_1 = A_2$}
Equations of motion for $\pi$ is given by
\begin{align}\label{eq:case3}
\mathcal{E}_4^{A_1 \neq A_2} &= A_1\,\left[\, \frac{3}{2}\,\Box \pi\, \partial_{\mu}\partial_{\nu}\pi^* \,\partial^{\mu} \partial^{ \nu}\pi^* + 3\,\Box \pi^*\, \partial_{\mu}\partial_{\nu}\pi \,\partial^{\mu} \partial^{ \nu}\pi^* -3\,\partial_{\mu}\partial_{\nu}\pi^*\, \partial^{\nu}\partial^{\alpha}\pi\,\partial^{\mu}\partial_{\alpha}\pi^*  -\frac{3}{2}\,(\Box \pi^*)^2\,\Box \pi\,\,\right]
\end{align}
which in the limit $\pi = \pi^*$ gives
\begin{align}\label{eq:case3_scalarlimit}
\mathcal{E}_4^{A_1 = A_2} &= \frac{3 A_1}{2}\,\,[\, 3\, \Box \pi\, \partial_{\mu}\partial_{\nu}\pi \,\partial^{\mu} \partial^{ \nu}\pi - 2\,\partial_{\mu}\partial_{\nu}\pi\, \partial^{\nu}\partial^{\alpha}\pi\,\partial^{\mu}\partial_{\alpha}\pi -(\Box \pi)^3\,\,]
\end{align}
\textbf{Case 4: $A_1 \neq A_2$}
Equations of motion for $\pi$ is given by
\begin{align}\label{eq:case4}
\mathcal{E}_4^{A_1 \neq A_2} &= \left(A_1 + \frac{A_2}{2}\right)[ \,\,\Box \pi\, \partial_{\mu}\partial_{\nu}\pi^*\, \partial^{\mu} \partial^{ \nu}\pi^* + 2\,\Box \pi^*\, \partial_{\mu}\partial_{\nu}\pi \,\partial^{\mu} \partial^{ \nu}\pi^*  -2\,\partial_{\mu}\partial_{\nu}\pi^*\, \partial^{\nu}\partial^{\alpha}\pi\,\partial^{\mu}\partial_{\alpha}\pi^*  -(\,\Box \pi^*)^2\,\Box \pi\,]
\end{align}
which in the limit $\pi = \pi^*$ gives
\begin{align}\label{eq:case4_scalarlimit}
\mathcal{E}_4^{A_1 \neq A_2} &= \left(A_1 + \frac{A_2}{2}\right)[\, 3\, \Box \pi\, \partial_{\mu}\partial_{\nu}\pi \,\partial^{\mu} \partial^{ \nu}\pi - 2\,\partial_{\mu}\partial_{\nu}\pi\, \partial^{\nu}\partial^{\alpha}\pi\,\partial^{\mu}\partial_{\alpha}\pi -(\Box \pi)^3\,\,]
\end{align}
From the above cases, it is interesting to see that equations of motion are independent of the values of $A_1$ and $A_2$ and the equations of motion of all the four cases are identical to Ref.~\cite{2009-Nicolis-PRD} in flat space-time. Hence, for simplicity, we set $A_1 = 0$ in the Lagrangian for the fourth order complex scalar Galileon Lagrangian.
%%%%%
\section{Gauge fixing}\label{GaugeAppend}

In this Appendix, we show that the Galileon complex scalar action \eqref{action_full} is indeed invariant under the Gauge transformation.
Replacing, $\partial_{\mu} \rightarrow {D}_{\mu} = \partial_{\mu} + \alpha A_{\mu} $ (where we fix $\alpha$ at the end of the calculation) implies 
\begin{align}\label{G1}
\partial_{\mu}\partial^{\nu}\pi \rightarrow {D}_{\mu}{D}^{\nu}\pi= (\partial_{\mu} + \alpha A_{\mu} )(\partial^{\nu} + \alpha A^{\nu})\pi
={\partial_{\mu}\,^{\nu}{\pi}+ \alpha  (\partial_{\mu}{A^{\nu}} \pi+ A^{\nu}    \partial_{\mu}{\pi}+ A_{\mu}    \partial^{\nu}{\pi})+ \alpha\alpha A_{\mu} A^{\nu}   \pi}
\end{align}
Under the local U(1) gauge transformation: 
$\pi \rightarrow \pi e^{-i e \theta(x)}$, above expression becomes
\begin{align}\label{G2}
{D}_{\mu}{D}^{\nu}(\pi e^{-i e \theta}) &=e^{-ie\theta}(\partial_{\mu}\,^{\nu}{\pi}+\alpha \partial_{\mu}{A^{\nu}} \pi+A^{\nu} \alpha \partial_{\mu}{\pi}+A_{\mu} \alpha \partial^{\nu}{\pi}+A_{\mu} A^{\nu} \alpha \alpha \pi\nonumber \\
&{}-ie (\partial^{\nu}{\pi} \partial_{\mu}{\theta}+\partial_{\mu}{\pi} \partial^{\nu}{\theta}+\partial_{\mu}\,^{\nu}{\theta} \pi+A^{\nu} \alpha \partial_{\mu}{\theta} \pi+A_{\mu} \alpha \partial^{\nu}{\theta} \pi)+ie ie \partial_{\mu}{\theta} \partial^{\nu}{\theta} \pi
\end{align}
Under the gauge transformation $A_{\mu} \rightarrow A_{\mu} + \partial_{\mu}\theta$, Eq.~(\ref{G2}) becomes,
\begin{align}\label{G3}
{D}_{\mu}{D}^{\nu}(\pi e^{-i e \theta})&\rightarrow e^{-ie\theta}[\partial_{\mu}\,^{\nu}{\pi}+\alpha( \partial_{\mu}{A^{\nu}} \pi + \partial_{\mu}{\partial^{\nu}\theta} \pi+ A^{\nu}  \partial_{\mu}{\pi}+ \partial^{\nu}\theta  \partial_{\mu}{\pi} +A_{\mu}  \partial^{\nu}{\pi}+   \partial_{\mu}\theta  \partial^{\nu}{\pi}\nonumber \\&{}+ \alpha  A_{\mu} A^{\nu}\pi + \alpha A_{\mu} \partial^{\nu}\theta \pi + \alpha A^{\nu} \partial_{\mu}\theta \pi + \alpha \partial_{\mu}\theta \partial^{\nu}\theta \pi)
-ie (\partial^{\nu}{\pi} \partial_{\mu}{\theta}+\partial_{\mu}{\pi} \partial^{\nu}{\theta}+\partial_{\mu}\,^{\nu}{\theta} \pi\nonumber \\&{}+\alpha A^{\nu} \partial_{\mu}{\theta} \pi +\alpha \partial^{\nu}\theta  \partial_{\mu}{\theta} \pi +\alpha A_{\mu} \partial^{\nu}{\theta} \pi + \alpha \partial_{\mu} \partial^{\nu}{\theta} \pi )+ie ie \partial_{\mu}{\theta} \partial^{\nu}{\theta} \pi]
\end{align}
For the ${D}_{\mu}{D}^{\nu}\pi$ to be invariant under the local gauge transformations, we set  $\alpha = ie$ in Eq.~(\ref{G3}). Hence, the action (\ref{action_CSG}) is invariant under the following 
simultaneous transformations:
\begin{align}\label{eq:gauge_trans}
A_{\mu} \rightarrow A_{\mu} + \partial_{\mu}\theta\,\,;\hspace{0.5cm}
\pi \rightarrow \pi e^{-ie\theta(x)} 
\end{align}
and the quantity ${D}_{\mu}{D}^{\nu}\pi$ will transform as
\begin{align}\label{eq:termTransform}
{D}_{\mu}{D}^{\nu}(\pi e^{-ie\theta})= (\partial_{\mu}\,^{\nu}{\pi}+\alpha \partial_{\mu}{A^{\nu}} \pi+A^{\nu} \alpha \partial_{\mu}{\pi}+A_{\mu} \alpha \partial^{\nu}{\pi}+A_{\mu} A^{\nu} \alpha \alpha \pi)e^{-ie\theta}
\end{align}
Similar analysis can be done for $\pi^* \rightarrow \pi^* e^{ie\theta(x)}$ also.
%%%%%
\section{Galileon scalar electrodynamics in curved space-time}
\label{app:C}

In this Appendix, we show that the action \eqref{action_m} leads to 
higher derivative equations of motion. We also show addition of the 
non-minimal term leads to second order equations of motion. The variation of the action (\ref{action_m}) with respect to $\pi$ yield the EOM:
\begin{align}
\mathcal{E}^{\rm{min}}_4 &= \frac{\omega}{2 \lambda_{\pi}^6}[\,\,2\,\nabla_{\mu}\nabla^{\mu}\pi^*\, \nabla_{\nu}\nabla_{\alpha}\pi\,\nabla^{\nu}\nabla^{\alpha}\pi^* -\nabla_{\mu}\nabla^{\mu}\pi^*\, \nabla_{\nu}\nabla^{\nu}\pi^*\, \nabla_{\alpha} \nabla^{\alpha}\pi - 2\,\nabla_{\mu}\nabla_{\nu}\pi^*\, \nabla^{\mu}\nabla^{\alpha}\pi^*\,\nabla^{\nu}\nabla_{\alpha}\pi\nonumber\\ &{}+\nabla_{\mu}\nabla^{\mu}\pi \,\nabla^{\mu}\nabla^{\nu}\pi^*\, \nabla_{\mu}\nabla_{\nu}\pi^* + 2\,\nabla_{\alpha}\pi\, \nabla^{\alpha}\pi^*\,(\,\nabla^{\nu}\nabla_{\nu}\nabla_{\mu}\nabla^{\mu}\pi^*  - \nabla^{\nu}\nabla^{\mu}\nabla_{\mu}\nabla_{\nu}\pi^*\, )\nonumber\\&{} + \nabla^{\alpha}\pi^*\, \nabla_{\nu}\pi^*\,(\,\nabla^{\mu}\nabla_{\mu}\nabla^{\nu}\nabla_{\alpha}\pi + \nabla^{\alpha}\nabla^{\nu}\nabla_{\mu}\nabla^{\mu}\pi\, )  - \nabla^{\alpha}\pi^*\, \nabla_{\mu}\pi^*\,(\,\nabla^{\nu}\nabla^{\mu}\nabla_{\nu} \nabla_{\alpha}\pi + \nabla_{\alpha}\nabla_{\nu}\nabla^{\mu}\nabla^{\nu}\pi\, )\nonumber\\&{} +3\,\nabla^{\nu}\nabla^{\alpha}\pi\, \nabla_{\alpha}\pi^*\,(\,\nabla_{\nu}\nabla_{\mu}\nabla^{\mu}\pi^* - \nabla_{\mu}\nabla^{\mu}\nabla_{\nu}\pi^*\,) + 2\,\nabla^{\nu}\nabla^{\alpha}\pi^*\, \nabla_{\alpha}\pi\,(\,\nabla_{\nu}\nabla_{\mu}\nabla^{\mu}\pi^* - \nabla_{\mu}\nabla^{\mu}\nabla_{\nu}\pi^*\,)\nonumber\\&{}
+ \nabla^{\mu}\nabla_{\mu}\pi\, \nabla^{\alpha}\pi^*\,(\,\nabla_{\nu}\nabla^{\nu}\nabla_{\alpha}\pi^* - \nabla_{\alpha}\nabla^{\nu}\nabla_{\nu}\pi^*\,)+ 2\,\nabla^{\mu}\nabla^{\nu}\pi^*\, \nabla^{\alpha}\pi^*\,(\nabla_{\mu}\nabla_{\nu}\nabla_{\alpha}\pi - \nabla_{\nu}\nabla_{\mu}\nabla_{\alpha}\pi\,)\nonumber\\&{}
+\nabla^{\nu}\nabla^{\alpha}\pi^*\, \nabla_{\alpha}\pi^*\,(\,\nabla_{\nu}\nabla_{\mu}\nabla^{\mu}\pi - \nabla_{\mu}\nabla^{\mu}\nabla_{\nu}\pi\,) + 2\,\nabla^{\mu}\nabla^{\nu}\pi\, \nabla^{\alpha}\pi^*\,(\,\nabla_{\alpha}\nabla_{\mu}\nabla_{\nu}\pi^* - \nabla_{\nu}\nabla_{\alpha}\nabla_{\mu}\pi^*\,)\nonumber\\&{}
+2\,\nabla^{\mu}\nabla^{\nu}\pi^*\, \nabla^{\alpha}\pi\,(\,\nabla_{\mu}\nabla_{\alpha}\nabla_{\nu}\pi^* - \nabla_{\nu}\nabla_{\mu}\nabla_{\alpha}\pi^*\,)\,\,]
\end{align} 
Using the following commutation properties of covariant derivatives:
\begin{align}\label{eq:commutation}
[\nabla_{\mu},\nabla_{\nu}]\nabla^{\alpha}\pi = R^{\alpha}_{\rho\mu\nu}\,\nabla^{\rho}\pi; \;
\nabla_{\nu}\nabla_{\mu}\nabla_{\alpha}\pi - \nabla_{\alpha}\nabla_{\mu}\nabla_{\nu}\pi = R^{\rho}_{\mu\alpha\nu}\nabla_{\rho}\pi; \;
\nabla_{\alpha}\nabla_{\nu}\nabla^{\nu}\pi - \nabla_{\nu}\nabla^{\nu}\nabla_{\alpha}\pi = -R_{\nu\alpha}\nabla^{\nu}\pi,
\end{align}
we get
\begin{align}
\mathcal{E}_4^{\rm{min}} = \frac{\omega}{2 \lambda_{\pi}^6}[\,-2\,\nabla_{\alpha}\pi\, \nabla^{\alpha}\pi^*\,\nabla^{\nu}\nabla^{\mu}\pi^*\,R_{\mu\nu} -\nabla_{\alpha}\pi\, \nabla^{\alpha}\pi^*\,\nabla^{\mu}\pi^*\,\nabla_{\mu}R 
+2\,\nabla^{\mu}\pi^*\,\nabla^{\alpha}\pi^*\,\nabla^{\nu}\nabla^{\rho}\pi\, R_{\rho\mu\alpha\nu} \nonumber\\ - \nabla^{\mu}\pi^*\,\nabla^{\alpha}\pi^*\,\nabla_{\mu}\nabla^{\nu}\pi\, R_{\nu\alpha} - \nabla^{\mu}\pi^*\,\nabla^{\alpha}\pi^*\,\nabla^{\rho}\pi\,\nabla_{\rho}R_{\alpha\mu} -3\,\nabla^{\alpha}\nabla^{\nu}\pi\, \nabla_{\alpha}\pi^* \,\nabla^{\mu}\pi^*\, R_{\mu\nu}\nonumber\\ -2\,\nabla^{\mu}\pi^*\,\nabla_{\alpha}\pi\,\nabla^{\nu}\nabla^{\alpha}\pi^*\,R_{\mu\nu} + \nabla^{\nu}\pi^*\,\nabla^{\alpha}\pi^*\,\nabla^{\mu}\nabla_{\mu}\pi\, R_{\nu\alpha} -2\,\nabla^{\nu}\nabla^{\alpha}\pi^*\,\nabla_{\alpha}\pi^*\,\nabla^{\mu}\pi\, R_{\mu\nu}\nonumber\\ +2\,\nabla_{\mu}\nabla^{\mu}\pi^*\, \nabla_{\nu}\nabla_{\alpha}\pi\,\nabla^{\nu}\nabla^{\alpha}\pi^* -\nabla_{\mu}\nabla^{\mu}\pi^*\, \nabla_{\nu}\nabla^{\nu}\pi^*\, \nabla_{\alpha} \nabla^{\alpha}\pi - 2\,\nabla_{\mu}\nabla_{\nu}\pi^*\, \nabla^{\mu}\nabla^{\alpha}\pi^*\nabla^{\nu}\nabla_{\alpha}\pi\nonumber\\ +\nabla_{\mu}\nabla^{\mu}\pi\, \nabla^{\mu}\nabla^{\nu}\pi^* \,\nabla_{\mu}\nabla_{\nu}\pi^*\,\,]
\end{align}
Note that the second and fifth terms in the above equation are higher derivative (of metric) terms. In order to remove these terms we need to add some non-minimal terms in the action in such a way that on varying the total action, all the derivatives of the Ricci tensor and Ricci scalar vanish. The equation of motion for the non-minimal action (\ref{non-min}) is given by:
\begin{align}
\mathcal{E}^{\rm{nm}}_4 &= \frac{\omega}{4 \lambda_{\pi}^6}[\,2\,\nabla_{\mu}\nabla^{\mu}\pi^*\, \nabla_{\nu}\pi^*\,\nabla^{\nu}\pi\, R 
+ 2\,\nabla_{\mu}\nabla_{\nu}\pi\, \nabla^{\mu}\pi^*\,\nabla^{\nu}\pi^*\, R  + 2\, \nabla_{\mu}\pi^* \,\nabla^{\mu}\pi\,\nabla^{\alpha}\pi^* \,\nabla_{\alpha}R\nonumber\\&{}
-\nabla_{\mu}\nabla^{\mu}\pi\, \nabla_{\nu}\pi^*\,\nabla^{\nu}\pi^*\, R+ 2\,\nabla_{\mu}\nabla^{\mu}\pi\, \nabla^{\nu}\pi^*\,\nabla^{\alpha}\pi^*\, R_{\nu\alpha} + 4\, \nabla^{\alpha}\nabla^{\mu}\pi^*\,\nabla_{\alpha}\pi\,\nabla^{\nu}\pi^*\, R_{\mu\nu}\nonumber\\&{} + 2\,\nabla^{\alpha}\pi\, \nabla^{\mu}\pi^*\, \nabla^{\nu}\pi^*\, \nabla_{\alpha} R_{\mu\nu} + 4\,\nabla^{\mu}\nabla_{\alpha}\pi^*\,\nabla^{\alpha}\pi^*\, \nabla^{\nu}\pi  R_{\mu\nu} + + 2\nabla_{\alpha}\pi^*\nabla^{\alpha}\pi^* \nabla^{\mu}\nabla^{\nu}\pi\, R_{\mu\nu}\,\,]
\end{align}
If we add the above non minimal action in the action (\ref{action_m}) then it will cancel all the higher order derivative terms in the EOM. So varying the action
${S}_4^{\rm{min}} + {S}^{\rm{nm}}_4$ with respect to $\pi$, we get 
\begin{align}\label{eq:S4eomCurved}
\mathcal{E}^{\prime} &= \frac{\omega}{2 \lambda_{\pi}^6}\,\,[\,\,2\,\Box\pi^*\, \nabla_{\nu}\nabla_{\alpha}\pi\,\nabla^{\nu}\nabla^{\alpha}\pi^* -(\,\Box\pi^*)^2\, \Box\pi +\Box\pi\, \nabla^{\mu}\nabla^{\nu}\pi^*\, \nabla_{\mu}\nabla_{\nu}\pi^* - 2\,\nabla_{\mu}\nabla_{\nu}\pi^*\, \nabla^{\mu}\nabla^{\alpha}\pi^*\,\nabla^{\nu}\nabla_{\alpha}\pi \nonumber\\ &{} \hspace{1cm} + \Box\pi^*\,\nabla_{\mu}\pi^*\,\nabla^{\mu}\pi\, R -\frac{1}{2}\Box\pi\,\nabla_{\alpha}\pi^*\,\nabla^{\alpha}\pi^* \,R + \nabla_{\alpha}\pi^*\,\nabla_{\mu}\pi^*\,\nabla^{\mu}\nabla^{\alpha}\pi \,R +2 \,\Box\pi\, \nabla^{\nu}\pi^*\,\nabla^{\mu}\pi^* \,R_{\nu\mu} \nonumber\\ &{} \hspace{1cm} +\nabla_{\alpha}\pi^*\, \nabla^{\alpha}\pi^*\,\nabla^{\nu}\nabla^{\mu}\pi \,R_{\mu\nu}  - 2\,\nabla_{\alpha}\pi\, \nabla^{\alpha}\pi^*\,\nabla^{\nu}\nabla^{\mu}\pi^*\,R_{\mu\nu} - \nabla^{\mu}\pi^*\,\nabla^{\alpha}\pi^*\,\nabla_{\mu}\nabla^{\nu}\pi\, R_{\nu\alpha} \nonumber\\ &{} \hspace{1cm}+ 2\, \nabla^{\mu}\pi^*\,\nabla^{\alpha}\pi^*\,\nabla^{\nu}\nabla^{\rho}\pi\, R_{\rho\mu\alpha\nu} -3\,\nabla^{\alpha}\nabla^{\nu}\pi\, \nabla_{\alpha}\pi^*\, \nabla^{\mu}\pi^*\,R_{\mu\nu}  ]
\end{align}
\section{Consistency check with real scalar galileon }\label{sec:consistencyRealGal}
In this section, we show systematically that our fourth order action ($S^{\rm{min}}_4 + S^{\rm{nm}}_4$ ) exactly matches with the results of Deffayet et al~\cite{Deffayet2009} in the $\pi = \pi^*$ limit.  Eq.~ (\ref{eq:S4pi}) can be written as,
\begin{align} \label{eq:Smin+Snm}
\left. {S}_4^{\rm{min}} + {S}_4^{\rm{nm}} \right|_{\pi = \pi^*}  &= A_2 \, \int d^4x\,\,\sqrt{-g}\,\, \left[\, (\Box\pi)^2 \,\nabla_{\alpha}\pi\, \nabla^{\alpha}\pi + \Box \pi\, \nabla_{\nu}\pi \, \nabla^{\nu}\nabla^{\alpha}\pi \,\nabla_{\alpha}\pi  \right.
\nonumber\\
& ~~ - \left. \,\nabla_{\mu}\nabla_{\nu}\pi\, \nabla^{\mu}\nabla^{\nu}\pi\, \nabla_{\alpha} \pi\, \nabla^{\alpha}\pi 
 - \nabla^{\mu}\nabla^{\nu}\pi\, \nabla_{\nu}\nabla_{\alpha}\pi\,\nabla_{\mu}\pi\, \nabla^{\alpha}\pi\, \right]  \\
&{} 
 + \frac{A_2}{4} \int d^4x\,\,\sqrt{-g}\,\,\nabla_{\alpha}\pi \nabla^{\alpha}\pi \nabla^{\mu}\pi\nabla^{\nu}\pi\,\,G_{\mu\nu} - \frac{3A_2}{4} \int d^4x\,\,\sqrt{-g}\,\, \nabla_{\alpha}\pi\nabla^{\alpha}\pi \nabla^{\mu}\pi \nabla^{\nu}\pi R_{\mu\nu} \nonumber
\end{align}
where $G_{\mu\nu} = R_{\mu\nu} - \frac{1}{2} \, g_{\mu\nu} \, R$. 
Concentrating on the last term in the above equation (\ref{eq:Smin+Snm}) and using the relation $(\nabla_{\mu}\nabla_{\nu}\nabla^{\nu}\pi - \nabla_{\nu}\nabla^{\nu}\nabla_{\mu}\pi)\,\nabla^{\mu}\pi = R_{\mu\nu}\nabla^{\mu}\nabla^{\nu}\pi$, ignoring the total derivative terms, we get:
\begin{align}\label{eq:residualTerm}
  -\frac{3A_2}{4} \nabla_{\alpha}\pi\nabla^{\alpha}\pi \nabla^{\mu}\pi \nabla^{\nu}\pi R_{\mu\nu} &= -\frac{3A_2}{4} \nabla_{\alpha}\pi\nabla^{\alpha}\pi \nabla_{\mu}\nabla^{\mu}\nabla_{\nu}\pi \nabla^{\nu}\pi + \frac{3A_2}{4} \nabla_{\alpha}\pi\nabla^{\alpha}\pi \nabla_{\nu}\nabla^{\mu}\nabla_{\mu}\pi \nabla^{\nu}\pi  \nonumber \\
  &{} = \frac{3A_2}{2} \nabla_{\mu}\nabla_{\alpha}\pi \nabla^{\mu}\nabla^{\nu}\pi \nabla_{\nu}\pi \nabla^{\alpha}\pi + \frac{3A_2}{4} \nabla_{\mu}\nabla_{\nu}\pi \nabla^{\mu}\nabla^{\mu}\pi \nabla^{\nu}\pi \nabla_{\alpha}\pi\nabla^{\alpha}\pi   \nonumber \\
  &{}\hspace{1cm} -\frac{3A_2}{2} (\Box \pi )^2\nabla_{\alpha}\pi\nabla^{\alpha}\pi - \frac{3A_2}{4} \Box \pi \nabla^{\mu}\nabla^{\nu}\pi \nabla_{\mu}\pi \nabla_{\nu}\pi \, .
\end{align}
Using Eqs.~(\ref{eq:Smin+Snm}) and (\ref{eq:residualTerm}), we get:
\begin{align} \label{eq:CovariantGal_action}
\left. {S}_4^{\rm{min}} + {S}_4^{\rm{nm}} \right|_{\pi = \pi^*}  &= \frac{A_2}{4} \, \int d^4x\,\,\sqrt{-g}\,\, \left[\, (\Box\pi)^2 \,\nabla_{\alpha}\pi\, \nabla^{\alpha}\pi + \Box \pi\, \nabla_{\nu}\pi \, \nabla^{\nu}\nabla^{\alpha}\pi \,\nabla_{\alpha}\pi  \right.
\nonumber\\
&{} \hspace{1.5cm} - \left. \,\nabla_{\mu}\nabla_{\nu}\pi\, \nabla^{\mu}\nabla^{\nu}\pi\, \nabla_{\alpha} \pi\, \nabla^{\alpha}\pi 
 - \nabla^{\mu}\nabla^{\nu}\pi\, \nabla_{\nu}\nabla_{\alpha}\pi\,\nabla_{\mu}\pi\, \nabla^{\alpha}\pi\, \right] \nonumber \\
&{} 
 + \frac{A_2}{4} \int d^4x\,\,\sqrt{-g}\,\,\nabla_{\alpha}\pi \nabla^{\alpha}\pi \nabla^{\mu}\pi\nabla^{\nu}\pi\,\,G_{\mu\nu} 
\end{align}
Thus, the above action (\ref{eq:CovariantGal_action}) matches with the action in Ref.~\cite{Deffayet2009}.
It is important to note that there is an overall sign difference between our result and obtained by Deffayet et al~\cite{Deffayet2009}. This is due to the signature convention of the metric and hence our definition of energy-momentum tensor i.e.  
$T_{\mu\nu} = \frac{2}{\sqrt{-g}}\frac{\delta S}{\delta g^{\mu\nu}}$, differs by an overall negative sing compared to Deffayet et al~\cite{Deffayet2009}.

\section{Galileon scalar electrodynamics in curved space time for $A_2 = 0$}\label{sec:A1_terms}
Before discussing the other special case $A_2 = 0$, we obtain the non-minimal coupling terms for the full action.
Consider the action corresponding to the full Lagrangian (\ref{LCGG}) and varying the action with respect to $\pi$ gives the equation of motion for the generic action in curved spacetime. As our main focus is to obtain the non-minimal action, we collect only the fourth order derivative terms in the field as we have done in Appendix (\ref{app:C}).
After collecting the fourth order terms we get,
\begin{align}\label{eq:app_LCGGminimal}
\mathcal{E}^{\rm{min}}_4 &= \frac{A_2}{2} \left[\,\,2\nabla_{\alpha}\pi\, \nabla^{\alpha}\pi^*\,(\,\nabla^{\nu}\nabla_{\nu}\nabla_{\mu}\nabla^{\mu}\pi^*  - \nabla^{\nu}\nabla^{\mu}\nabla_{\mu}\nabla_{\nu}\pi^*\, ) \right. \nonumber\\&{} \left. + \nabla_{\alpha}\pi^*\, \nabla^{\mu}\pi^*\,(\,\nabla_{\nu}\nabla^{\nu}\nabla^{\alpha}\nabla_{\mu}\pi + \nabla_{\mu}\nabla^{\alpha}\nabla_{\nu}\nabla^{\nu}\pi  - \nabla_{\nu}\nabla^{\alpha}\nabla^{\nu} \nabla_{\mu}\pi - \nabla_{\mu}\nabla_{\nu}\nabla^{\alpha}\nabla^{\nu}\pi\, )\right] \nonumber \\&{} 
+ A_1 \left[\,\,2 \nabla_{\alpha}\pi^*\, \nabla^{\alpha}\pi^*\,(\,\nabla^{\nu}\nabla_{\nu}\nabla_{\mu}\nabla^{\mu}\pi  - \nabla^{\nu}\nabla^{\mu}\nabla_{\mu}\nabla_{\nu}\pi\, ) \right. \\
%%%%%
&{} \left. + \nabla^{\alpha}\pi^*\, \nabla^{\mu}\pi^*\,(\,\nabla_{\nu}\nabla^{\alpha}\nabla^{\nu}\nabla_{\mu}\pi + \nabla_{\mu}\nabla_{\nu}\nabla^{\alpha}\nabla^{\nu}\pi  - \nabla^{\nu}\nabla_{\nu}\nabla^{\alpha} \nabla_{\mu}\pi - \nabla_{\mu}\nabla^{\alpha}\nabla_{\nu}\nabla^{\nu}\pi\, ) \right. \nonumber\\&{} \left. + 2 \nabla_{\alpha}\pi^*\, \nabla_{\nu}\pi\,(\,\nabla^{\mu}\nabla_{\mu}\nabla^{\nu}\nabla^{\alpha}\pi^* + \nabla^{\nu}\nabla^{\alpha}\nabla^{\mu}\nabla_{\mu}\pi^*  - \nabla^{\nu}\nabla^{\mu}\nabla^{\alpha} \nabla_{\mu}\pi^* - \nabla_{\mu}\nabla_{\alpha}\nabla^{\mu}\nabla_{\nu}\pi^*\, ) \right]
\nonumber 
\end{align} 
using the commutation properties of the covariant derivatives and using (\ref{eq:commutation}) we get
\begin{align} \label{eq:higherordertermsLCGG}
\mathcal{E}^{\rm{min}}_4 &= -\frac{A_2}{2} \nabla_{\alpha}\pi^* \nabla^{\alpha}\pi \nabla^{\mu}\pi^* \nabla_{\mu}R - A_1\nabla_{\alpha}\pi^* \nabla^{\alpha}\pi^* \nabla^{\mu}\pi \nabla_{\mu}R  -\frac{A_2}{2} \nabla^{\alpha}\pi^* \nabla^{\mu}\pi^* \nabla^{\nu}\pi \nabla_{\nu}R_{\alpha\mu} \nonumber\\ &{} + A_1 \nabla^{\alpha}\pi^* \nabla^{\mu}\pi^* \nabla^{\nu}\pi \nabla_{\nu}R_{\alpha\mu} - 2A_1 \nabla^{\alpha}\pi^* \nabla^{\mu}\pi^* \nabla^{\nu}\pi \nabla_{\mu}R_{\alpha\nu} \, .
\end{align} 
We now consider the following general non-minimal action
\begin{align}\label{eq:general_nm_action}
    S^{\rm{nm}}_4 = \int d^4x \sqrt{-g} \,\, \mathcal{L}^{\rm{nm}}_4
\end{align}
where 
\begin{align}\label{eq:general_nm_lagrangian}
 \mathcal{L}^{\rm{nm}}_4 &= - \left[ \nabla_{\alpha}\pi^* \nabla^{\alpha}\pi^* \nabla^{\mu}\pi \nabla^{\nu}\pi + \nabla_{\alpha}\pi \nabla^{\alpha}\pi \nabla^{\mu}\pi^* \nabla^{\nu}\pi^* \right]\left(g_1 \, R_{\mu\nu} + g_2 \, g_{\mu\nu} R \right) \nonumber\\&{} -\nabla_{\alpha}\pi^* \nabla^{\alpha}\pi \nabla^{\mu}\pi^* \nabla^{\nu}\pi ]\left(h_1 \, R_{\mu\nu} + h_2 \, g_{\mu\nu} R \right)
\end{align}
where $g_1, g_2, h_1$ and $h_2$ are arbitray constants which we will fix later.
Variation of the action (\ref{eq:general_nm_action}) with respect to $\pi$ and collecting only derivatives of the Ricci tensor and Ricci scalar terms because we are interested in fixing the coefficients  $g_1, g_2, h_1,h_2$, we obtain
\begin{align}\label{eq:higherorderterms_nm}
\mathcal{E}^{\rm{nm}}_4 &= g_1 \nabla_{\alpha}\pi^* \nabla^{\alpha}\pi^* \nabla^{\mu}\pi \nabla_{\mu}R + 4g_2 \nabla_{\alpha}\pi^* \nabla^{\alpha}\pi^* \nabla^{\mu}\pi \nabla_{\mu}R + \frac{h_1}{2} \nabla_{\alpha}\pi^* \nabla^{\alpha}\pi \nabla^{\mu}\pi^* \nabla_{\mu}R \\ &{} + 2h_2 \nabla_{\alpha}\pi^* \nabla^{\alpha}\pi \nabla^{\mu}\pi^* \nabla_{\mu}R + 2g_1 \nabla^{\mu}\pi^* \nabla^{\nu}\pi^* \nabla^{\alpha}\pi \nabla_{\alpha}R_{\mu\nu} + h_1 \nabla^{\mu}\pi^* \nabla^{\nu}\pi  \nabla^{\alpha}\pi^* \nabla_{\alpha}R_{\mu\nu} \, .
\nonumber
\end{align} 
Adding Eqs.~(\ref{eq:higherordertermsLCGG}) and (\ref{eq:higherorderterms_nm}) such that all the derivative terms of Ricci tensor and scalar vanish, we obtain the following relations
:
\begin{align}
    -A_1 + g_1 + 4g_2 &= 0 \\
    -\frac{A_2}{2} + A_1 + 2g_1 &= 0\\
    -\frac{A_2}{2} + 2h_2 + \frac{h_1}{2} &= 0\\
    -2A_1 + h_1 &= 0
\end{align}
which implies 
\begin{align}
    g_1 &=  \frac{A_2}{4}  -\frac{A_1}{2} \\
    g_2 &= \frac{3A_1}{8} -\frac{A_2}{16}\\
    h_1 &= 2A_1\\
    h_2 &= -\frac{A_1}{2} + \frac{A_2}{4} \, .
\end{align}
Substituting the above values in Eq.~(\ref{eq:general_nm_lagrangian}), we get
\begin{align}\label{eq:general_nm_lagrangianA1A2}
 \mathcal{L}^{\rm{nm}}_4 &= \frac{A_1}{2}\left[ \nabla_{\alpha}\pi^* \nabla^{\alpha}\pi^* \nabla^{\mu}\pi \nabla^{\nu}\pi + \nabla_{\alpha}\pi \nabla^{\alpha}\pi \nabla^{\mu}\pi^* \nabla^{\nu}\pi^* \right]\left( R_{\mu\nu} - \frac{3}{4} \, g_{\mu\nu} R \right) \nonumber\\&{}
- \frac{A_2}{4}\left[ \nabla_{\alpha}\pi^* \nabla^{\alpha}\pi^* \nabla^{\mu}\pi \nabla^{\nu}\pi + \nabla_{\alpha}\pi \nabla^{\alpha}\pi \nabla^{\mu}\pi^* \nabla^{\nu}\pi^* \right]\left( R_{\mu\nu} - \frac{1}{4} \, g_{\mu\nu} R \right) \nonumber\\&{} -2A_1\nabla_{\alpha}\pi^* \nabla^{\alpha}\pi \nabla^{\mu}\pi^* \nabla^{\nu}\pi ]\left( R_{\mu\nu} - \frac{1}{4}\, g_{\mu\nu} R \right) - \frac{A_2}{4}\nabla_{\alpha}\pi^* \nabla^{\alpha}\pi \nabla^{\mu}\pi^* \nabla^{\nu}\pi \, g_{\mu\nu} R
\end{align}
We note that the above non-minimal action identical with $A_1 = 0$ case considered in Appendix~(\ref{app:C}). In $\pi = \pi^*$ limit, this case also leads to identical equations of motion as in Ref.~\cite{Deffayet2009}.

\section{Vector Galileon action $S_{VEG}$}
\label{app:D}

In Ref.~\cite{2017-Nandi.Shankaranarayanan-JCAP}, the authors obtained a vector Galileon model that leads to second-order equations. 
For completeness, in this appendix, we have listed below the vector Galileon action. The complete vector Galileon action can be written as
\begin{equation}\label{eq:VEG}
{S}_{VEC} = {S}_{\rm VG} + \lambda_{\rm VG} \sum_{i=1}^{12} {S}_{Vi}.
\end{equation}
where, $\lambda_{VG}$ is coupling constant 
\begin{align}
\label{eq:VecGalAction}
{S}_{VG} &= \lambda_{\rm VG}\,\int \,d^4 x\, \sqrt{-g}\,
\epsilon^{\alpha \gamma \nu} \epsilon^{\mu \eta \beta}\,
\nabla_{\alpha \beta}A_\gamma \, \nabla_{\mu \nu}A_\eta \\
{S}_{V1} &= E_1\,\int d^4x\, \sqrt{-g}\, {g}^{\mu \nu}
{g}^{\alpha \beta} {g}^{\gamma \delta}\, R_{\mu \nu}
\,{\nabla}_{\alpha}{A}_{\gamma}\, {\nabla}_{\beta}{A}_{\delta}
\\ {S}_{V2} &= E_2\,\int d^4x\, \sqrt{-g}\,{g}^{\mu \alpha} {g}^{\nu \beta} {g}^{\gamma \delta}\,{R}_{\mu
	\nu}\, {\nabla}_{\alpha}{A}_{\gamma}\,
{\nabla}_{\beta}{A}_{\delta} \\
{S}_{V3} &= E_3\,\int d^4x\, \sqrt{-g}\,{g}^{\mu \nu} {g}^{\alpha \beta}
{g}^{\gamma \delta} \,{R}_{\mu \nu}
\,{\nabla}_{\alpha}{A}_{\beta}
\,{\nabla}_{\gamma}{A}_{\delta} \\ 
{S}_{V4} &=E_4\,\int d^4x\, \sqrt{-g}\,{g}^{\mu \nu} {g}^{\alpha
	\delta} {g}^{\gamma \beta}\, {R}_{\mu \nu}
\,{\nabla}_{\alpha}{A}_{\beta}\,
{\nabla}_{\gamma}{A}_{\delta} \\ 
{S}_{V5} &=E_5\,\int d^4x\, \sqrt{-g}\,{g}^{\mu \gamma} {g}^{\alpha
	\beta} {g}^{\nu \delta}\, {R}_{\mu \nu}\,
{\nabla}_{\alpha}{A}_{\beta}\,
{\nabla}_{\gamma}{A}_{\delta} \\ 
{S}_{V6} &=E_{6}\,\int d^4x\, \sqrt{-g}\,{g}^{\mu \alpha} {g}^{\nu
	\delta} {g}^{\gamma \beta}\, {R}_{\mu \nu}\,
{\nabla}_{\alpha}{A}_{\beta}\,
{\nabla}_{\gamma}{A}_{\delta} \\
{S}_{V7} &=E_{7}\,\int d^4x\, \sqrt{-g}\,{g}^{\mu \alpha} {g}^{\nu
	\beta} {g}^{\gamma \zeta} {g}^{\delta \eta}
\,{R}_{\alpha \beta \gamma \delta}\,
{\nabla}_{\mu}{A}_{\nu}\, {\nabla}_{\zeta}{A}_{\eta}\\
{S}_{V8}&= E_{8}\,\int d^4x\,
\sqrt{-g}\,{g}^{\mu \alpha} {g}^{\eta \beta}
{g}^{\gamma \zeta} {g}^{\delta \nu} \,{R}_{\alpha
	\beta \gamma \delta}\, {\nabla}_{\mu}{A}_{\nu}
\,{\nabla}_{\zeta}{A}_{\eta}~~~~~~
\\ 
{S}_{V9} &= E_{9}\,\int d^4x\,\sqrt{-g}\,{g}^{\alpha \beta} {g}^{\gamma \delta}
{g}^{\mu \nu}\, {R}_{\alpha \beta}\,{R}_{\gamma \delta}\, {A}_{\mu}{A}_{\nu}\\ 
{S}_{V10} &= E_{10}\,\int d^4x\, \sqrt{-g}\,{g}^{\alpha \beta}{g}^{\gamma \mu} {g}^{\delta \nu}\,
{R}_{\alpha \beta}\, {R}_{\gamma \delta}\,{A}_{\mu} {A}_{\nu} \\ 
{S}_{V11}&=
E_{11}\,\int d^4x\, \sqrt{-g}\,{g}^{\alpha\gamma} {g}^{\beta \delta} {g}^{\mu \nu}\,
{R}_{\alpha \beta}\, {R}_{\gamma \delta}\,{A}_{\mu} {A}_{\nu}~~~~~~
\\ {S}_{V12} &= E_{12}\,\int d^4x\,\sqrt{-g}\,{g}^{\alpha \gamma} {g}^{\beta
	\mu} {g}^{\delta \nu}\, {R}_{\alpha\beta}\, {R}_{\gamma \delta}\, {A}_{\mu}{A}_{\nu}
\end{align} 
where $E_i$'s are dimensionless coefficients. Note that $E_i$ are related to $\lambda_{\rm VG}$. 
%\bibliography{References.bib}

%	\bibliography{masterref}

\begin{thebibliography}{31}%
\makeatletter
\providecommand \@ifxundefined [1]{%
 \@ifx{#1\undefined}
}%
\providecommand \@ifnum [1]{%
 \ifnum #1\expandafter \@firstoftwo
 \else \expandafter \@secondoftwo
 \fi
}%
\providecommand \@ifx [1]{%
 \ifx #1\expandafter \@firstoftwo
 \else \expandafter \@secondoftwo
 \fi
}%
\providecommand \natexlab [1]{#1}%
\providecommand \enquote  [1]{``#1''}%
\providecommand \bibnamefont  [1]{#1}%
\providecommand \bibfnamefont [1]{#1}%
\providecommand \citenamefont [1]{#1}%
\providecommand \href@noop [0]{\@secondoftwo}%
\providecommand \href [0]{\begingroup \@sanitize@url \@href}%
\providecommand \@href[1]{\@@startlink{#1}\@@href}%
\providecommand \@@href[1]{\endgroup#1\@@endlink}%
\providecommand \@sanitize@url [0]{\catcode `\\12\catcode `\$12\catcode
  `\&12\catcode `\#12\catcode `\^12\catcode `\_12\catcode `\%12\relax}%
\providecommand \@@startlink[1]{}%
\providecommand \@@endlink[0]{}%
\providecommand \url  [0]{\begingroup\@sanitize@url \@url }%
\providecommand \@url [1]{\endgroup\@href {#1}{\urlprefix }}%
\providecommand \urlprefix  [0]{URL }%
\providecommand \Eprint [0]{\href }%
\providecommand \doibase [0]{http://dx.doi.org/}%
\providecommand \selectlanguage [0]{\@gobble}%
\providecommand \bibinfo  [0]{\@secondoftwo}%
\providecommand \bibfield  [0]{\@secondoftwo}%
\providecommand \translation [1]{[#1]}%
\providecommand \BibitemOpen [0]{}%
\providecommand \bibitemStop [0]{}%
\providecommand \bibitemNoStop [0]{.\EOS\space}%
\providecommand \EOS [0]{\spacefactor3000\relax}%
\providecommand \BibitemShut  [1]{\csname bibitem#1\endcsname}%
\let\auto@bib@innerbib\@empty
%</preamble>
\bibitem [{\citenamefont {Podolsky}\ and\ \citenamefont
  {Schwed}(1948)}]{1948-Podolsky.Schwed-RMP}%
  \BibitemOpen
  \bibfield  {author} {\bibinfo {author} {\bibfnamefont {B.}~\bibnamefont
  {Podolsky}}\ and\ \bibinfo {author} {\bibfnamefont {P.}~\bibnamefont
  {Schwed}},\ }\href {\doibase 10.1103/RevModPhys.20.40} {\bibfield  {journal}
  {\bibinfo  {journal} {Rev. Mod. Phys.}\ }\textbf {\bibinfo {volume} {20}},\
  \bibinfo {pages} {40} (\bibinfo {year} {1948})}\BibitemShut {NoStop}%
\bibitem [{\citenamefont {{Thirring}}(1950)}]{1950-Thirring-PR}%
  \BibitemOpen
  \bibfield  {author} {\bibinfo {author} {\bibfnamefont {W.}~\bibnamefont
  {{Thirring}}},\ }\href {\doibase 10.1103/PhysRev.79.703} {\bibfield
  {journal} {\bibinfo  {journal} {Physical Review}\ }\textbf {\bibinfo {volume}
  {79}},\ \bibinfo {pages} {703} (\bibinfo {year} {1950})}\BibitemShut
  {NoStop}%
\bibitem [{\citenamefont {Pais}\ and\ \citenamefont
  {Uhlenbeck}(1950)}]{1950-Pais.Uhlenbeck-PR}%
  \BibitemOpen
  \bibfield  {author} {\bibinfo {author} {\bibfnamefont {A.}~\bibnamefont
  {Pais}}\ and\ \bibinfo {author} {\bibfnamefont {G.~E.}\ \bibnamefont
  {Uhlenbeck}},\ }\href {\doibase 10.1103/PhysRev.79.145} {\bibfield  {journal}
  {\bibinfo  {journal} {Phys. Rev.}\ }\textbf {\bibinfo {volume} {79}},\
  \bibinfo {pages} {145} (\bibinfo {year} {1950})}\BibitemShut {NoStop}%
%%CITATION = PHRVA,79,145;%%
\bibitem [{\citenamefont {Woodard}(2015)}]{2015-Woodard-arXiv}%
  \BibitemOpen
  \bibfield  {author} {\bibinfo {author} {\bibfnamefont {R.~P.}\ \bibnamefont
  {Woodard}},\ }\href {\doibase 10.4249/scholarpedia.32243} {\bibfield
  {journal} {\bibinfo  {journal} {Scholarpedia}\ }\textbf {\bibinfo {volume}
  {10}},\ \bibinfo {pages} {32243} (\bibinfo {year} {2015})},\ \Eprint
  {http://arxiv.org/abs/1506.02210} {arXiv:1506.02210 [hep-th]} \BibitemShut
  {NoStop}%
%%CITATION = ARXIV:1506.02210;%%
\bibitem [{\citenamefont {Simon}(1990)}]{1990-Simon-PRD}%
  \BibitemOpen
  \bibfield  {author} {\bibinfo {author} {\bibfnamefont {J.~Z.}\ \bibnamefont
  {Simon}},\ }\href {\doibase 10.1103/PhysRevD.41.3720} {\bibfield  {journal}
  {\bibinfo  {journal} {Phys. Rev. D.}\ }\textbf {\bibinfo {volume} {D41}},\
  \bibinfo {pages} {3720} (\bibinfo {year} {1990})}\BibitemShut {NoStop}%
%%CITATION = PHRVA,D41,3720;%%
\bibitem [{\citenamefont {Hawking}\ and\ \citenamefont
  {Hertog}(2002)}]{2002-Hawking.Hertog-PRD}%
  \BibitemOpen
  \bibfield  {author} {\bibinfo {author} {\bibfnamefont {S.~W.}\ \bibnamefont
  {Hawking}}\ and\ \bibinfo {author} {\bibfnamefont {T.}~\bibnamefont
  {Hertog}},\ }\href {\doibase 10.1103/PhysRevD.65.103515} {\bibfield
  {journal} {\bibinfo  {journal} {Phys. Rev.}\ }\textbf {\bibinfo {volume}
  {D65}},\ \bibinfo {pages} {103515} (\bibinfo {year} {2002})},\ \Eprint
  {http://arxiv.org/abs/hep-th/0107088} {arXiv:hep-th/0107088} \BibitemShut
  {NoStop}%
%%CITATION = HEP-TH/0107088;%%
\bibitem [{\citenamefont {Barth}\ and\ \citenamefont
  {Christensen}(1983)}]{1983-Barth.Christensen-PRD}%
  \BibitemOpen
  \bibfield  {author} {\bibinfo {author} {\bibfnamefont {N.~H.}\ \bibnamefont
  {Barth}}\ and\ \bibinfo {author} {\bibfnamefont {S.~M.}\ \bibnamefont
  {Christensen}},\ }\href {\doibase 10.1103/PhysRevD.28.1876} {\bibfield
  {journal} {\bibinfo  {journal} {Phys. Rev. D.}\ }\textbf {\bibinfo {volume}
  {28}},\ \bibinfo {pages} {1876} (\bibinfo {year} {1983})}\BibitemShut
  {NoStop}%
%%CITATION = PHRVA,D28,1876;%%
\bibitem [{\citenamefont {Stelle}(1977)}]{1977-Stelle-PRD}%
  \BibitemOpen
  \bibfield  {author} {\bibinfo {author} {\bibfnamefont {K.~S.}\ \bibnamefont
  {Stelle}},\ }\href {\doibase 10.1103/PhysRevD.16.953} {\bibfield  {journal}
  {\bibinfo  {journal} {Phys. Rev.}\ }\textbf {\bibinfo {volume} {D16}},\
  \bibinfo {pages} {953} (\bibinfo {year} {1977})}\BibitemShut {NoStop}%
%%CITATION = PHRVA,D16,953;%%
\bibitem [{\citenamefont {Fradkin}\ and\ \citenamefont
  {Tseytlin}(1982)}]{1982-Fradkin.Tseytlin-NPB}%
  \BibitemOpen
  \bibfield  {author} {\bibinfo {author} {\bibfnamefont {E.~S.}\ \bibnamefont
  {Fradkin}}\ and\ \bibinfo {author} {\bibfnamefont {A.~A.}\ \bibnamefont
  {Tseytlin}},\ }\href {\doibase 10.1016/0550-3213(82)90444-8} {\bibfield
  {journal} {\bibinfo  {journal} {Nucl. Phys.}\ }\textbf {\bibinfo {volume}
  {B201}},\ \bibinfo {pages} {469} (\bibinfo {year} {1982})}\BibitemShut
  {NoStop}%
%%CITATION = NUPHA,B201,469;%%
\bibitem [{\citenamefont {Sotiriou}\ and\ \citenamefont
  {Faraoni}(2010)}]{2010-Sotiriou.Faraoni-RMP}%
  \BibitemOpen
  \bibfield  {author} {\bibinfo {author} {\bibfnamefont {T.~P.}\ \bibnamefont
  {Sotiriou}}\ and\ \bibinfo {author} {\bibfnamefont {V.}~\bibnamefont
  {Faraoni}},\ }\href {\doibase 10.1103/RevModPhys.82.451} {\bibfield
  {journal} {\bibinfo  {journal} {Rev. Mod. Phys.}\ }\textbf {\bibinfo {volume}
  {82}},\ \bibinfo {pages} {451} (\bibinfo {year} {2010})},\ \Eprint
  {http://arxiv.org/abs/0805.1726} {arXiv:0805.1726 [gr-qc]} \BibitemShut
  {NoStop}%
%%CITATION = ARXIV:0805.1726;%%
\bibitem [{\citenamefont {Capozziello}\ and\ \citenamefont
  {De~Laurentis}(2011)}]{2011-Capozziello.DeLaurentis-PRep}%
  \BibitemOpen
  \bibfield  {author} {\bibinfo {author} {\bibfnamefont {S.}~\bibnamefont
  {Capozziello}}\ and\ \bibinfo {author} {\bibfnamefont {M.}~\bibnamefont
  {De~Laurentis}},\ }\href {\doibase 10.1016/j.physrep.2011.09.003} {\bibfield
  {journal} {\bibinfo  {journal} {Phys. Rept.}\ }\textbf {\bibinfo {volume}
  {509}},\ \bibinfo {pages} {167} (\bibinfo {year} {2011})},\ \Eprint
  {http://arxiv.org/abs/1108.6266} {arXiv:1108.6266 [gr-qc]} \BibitemShut
  {NoStop}%
%%CITATION = ARXIV:1108.6266;%%
\bibitem [{\citenamefont {Nojiri}\ and\ \citenamefont
  {Odintsov}(2011)}]{2011-Nojiri.Odintsov-PRep}%
  \BibitemOpen
  \bibfield  {author} {\bibinfo {author} {\bibfnamefont {S.}~\bibnamefont
  {Nojiri}}\ and\ \bibinfo {author} {\bibfnamefont {S.~D.}\ \bibnamefont
  {Odintsov}},\ }\href {\doibase 10.1016/j.physrep.2011.04.001} {\bibfield
  {journal} {\bibinfo  {journal} {Phys. Rept.}\ }\textbf {\bibinfo {volume}
  {505}},\ \bibinfo {pages} {59} (\bibinfo {year} {2011})},\ \Eprint
  {http://arxiv.org/abs/1011.0544} {arXiv:1011.0544 [gr-qc]} \BibitemShut
  {NoStop}%
%%CITATION = ARXIV:1011.0544;%%
\bibitem [{\citenamefont {Clifton}\ \emph {et~al.}(2012)\citenamefont
  {Clifton}, \citenamefont {Ferreira}, \citenamefont {Padilla},\ and\
  \citenamefont {Skordis}}]{2012-Clifton.etal-PRep}%
  \BibitemOpen
  \bibfield  {author} {\bibinfo {author} {\bibfnamefont {T.}~\bibnamefont
  {Clifton}}, \bibinfo {author} {\bibfnamefont {P.~G.}\ \bibnamefont
  {Ferreira}}, \bibinfo {author} {\bibfnamefont {A.}~\bibnamefont {Padilla}}, \
  and\ \bibinfo {author} {\bibfnamefont {C.}~\bibnamefont {Skordis}},\ }\href
  {\doibase 10.1016/j.physrep.2012.01.001} {\bibfield  {journal} {\bibinfo
  {journal} {Phys. Rept.}\ }\textbf {\bibinfo {volume} {513}},\ \bibinfo
  {pages} {1} (\bibinfo {year} {2012})},\ \Eprint
  {http://arxiv.org/abs/1106.2476} {arXiv:1106.2476 [astro-ph.CO]} \BibitemShut
  {NoStop}%
%%CITATION = ARXIV:1106.2476;%%
\bibitem [{\citenamefont {Nojiri}\ \emph {et~al.}(2017)\citenamefont {Nojiri},
  \citenamefont {Odintsov},\ and\ \citenamefont
  {Oikonomou}}]{2017-Nojiri.etal-PRep}%
  \BibitemOpen
  \bibfield  {author} {\bibinfo {author} {\bibfnamefont {S.}~\bibnamefont
  {Nojiri}}, \bibinfo {author} {\bibfnamefont {S.~D.}\ \bibnamefont
  {Odintsov}}, \ and\ \bibinfo {author} {\bibfnamefont {V.~K.}\ \bibnamefont
  {Oikonomou}},\ }\href {\doibase 10.1016/j.physrep.2017.06.001} {\bibfield
  {journal} {\bibinfo  {journal} {Phys. Rept.}\ }\textbf {\bibinfo {volume}
  {692}},\ \bibinfo {pages} {1} (\bibinfo {year} {2017})},\ \Eprint
  {http://arxiv.org/abs/1705.11098} {arXiv:1705.11098} \BibitemShut {NoStop}%
%%CITATION = ARXIV:1705.11098;%%
\bibitem [{\citenamefont {Ishak}(2019)}]{Ishak:2018his}%
  \BibitemOpen
  \bibfield  {author} {\bibinfo {author} {\bibfnamefont {M.}~\bibnamefont
  {Ishak}},\ }\href {\doibase 10.1007/s41114-018-0017-4} {\bibfield  {journal}
  {\bibinfo  {journal} {Living Rev. Rel.}\ }\textbf {\bibinfo {volume} {22}},\
  \bibinfo {pages} {1} (\bibinfo {year} {2019})},\ \Eprint
  {http://arxiv.org/abs/1806.10122} {arXiv:1806.10122 [astro-ph.CO]}
  \BibitemShut {NoStop}%
%%CITATION = ARXIV:1806.10122;%%
\bibitem [{\citenamefont {Woodard}(2007)}]{2007-Woodard-Proc}%
  \BibitemOpen
  \bibfield  {author} {\bibinfo {author} {\bibfnamefont {R.~P.}\ \bibnamefont
  {Woodard}},\ }in\ \href {\doibase 10.1007/978-3-540-71013-4_14} {\emph
  {\bibinfo {booktitle} {3rd Aegean Summer School: The Invisible Universe: Dark
  Matter and Dark Energy}}},\ Vol.\ \bibinfo {volume} {720}\ (\bibinfo {year}
  {2007})\ pp.\ \bibinfo {pages} {403--433},\ \Eprint
  {http://arxiv.org/abs/astro-ph/0601672} {arXiv:astro-ph/0601672 [astro-ph]}
  \BibitemShut {NoStop}%
%%CITATION = ASTRO-PH/0601672;%%
\bibitem [{\citenamefont {Horndeski}(1974)}]{1974-Horndeski-IJTP}%
  \BibitemOpen
  \bibfield  {author} {\bibinfo {author} {\bibfnamefont {G.~W.}\ \bibnamefont
  {Horndeski}},\ }\href {\doibase 10.1007/BF01807638} {\bibfield  {journal}
  {\bibinfo  {journal} {Int. J. Theor. Phys.}\ }\textbf {\bibinfo {volume}
  {10}},\ \bibinfo {pages} {363} (\bibinfo {year} {1974})}\BibitemShut
  {NoStop}%
\bibitem [{\citenamefont {Deffayet}\ \emph
  {et~al.}(2009{\natexlab{a}})\citenamefont {Deffayet}, \citenamefont {Deser},\
  and\ \citenamefont {Esposito-Farese}}]{2009-Deffayet.etal4-PRD}%
  \BibitemOpen
  \bibfield  {author} {\bibinfo {author} {\bibfnamefont {C.}~\bibnamefont
  {Deffayet}}, \bibinfo {author} {\bibfnamefont {S.}~\bibnamefont {Deser}}, \
  and\ \bibinfo {author} {\bibfnamefont {G.}~\bibnamefont {Esposito-Farese}},\
  }\href {\doibase 10.1103/PhysRevD.80.064015} {\bibfield  {journal} {\bibinfo
  {journal} {Phys. Rev.}\ }\textbf {\bibinfo {volume} {D80}},\ \bibinfo {pages}
  {064015} (\bibinfo {year} {2009}{\natexlab{a}})},\ \Eprint
  {http://arxiv.org/abs/0906.1967} {arXiv:0906.1967 [gr-qc]} \BibitemShut
  {NoStop}%
%%CITATION = ARXIV:0906.1967;%%
\bibitem [{\citenamefont {Deffayet}\ \emph {et~al.}(2014)\citenamefont
  {Deffayet}, \citenamefont {G{\"u}mr{\"u}k{\c{c}}{\"u}o{\u{g}}lu},
  \citenamefont {Mukohyama},\ and\ \citenamefont
  {Wang}}]{2014-Deffayet.etal-JHEP}%
  \BibitemOpen
  \bibfield  {author} {\bibinfo {author} {\bibfnamefont {C.}~\bibnamefont
  {Deffayet}}, \bibinfo {author} {\bibfnamefont {A.~E.}\ \bibnamefont
  {G{\"u}mr{\"u}k{\c{c}}{\"u}o{\u{g}}lu}}, \bibinfo {author} {\bibfnamefont
  {S.}~\bibnamefont {Mukohyama}}, \ and\ \bibinfo {author} {\bibfnamefont
  {Y.}~\bibnamefont {Wang}},\ }\href {\doibase 10.1007/JHEP04(2014)082}
  {\bibfield  {journal} {\bibinfo  {journal} {Journal of High Energy Physics}\
  }\textbf {\bibinfo {volume} {2014}},\ \bibinfo {pages} {82} (\bibinfo {year}
  {2014})}\BibitemShut {NoStop}%
\bibitem [{\citenamefont {Lovelock}(1971)}]{1971-Lovelock-JMP}%
  \BibitemOpen
  \bibfield  {author} {\bibinfo {author} {\bibfnamefont {D.}~\bibnamefont
  {Lovelock}},\ }\href {\doibase 10.1063/1.1665613} {\bibfield  {journal}
  {\bibinfo  {journal} {J. Math. Phys.}\ }\textbf {\bibinfo {volume} {12}},\
  \bibinfo {pages} {498} (\bibinfo {year} {1971})}\BibitemShut {NoStop}%
%%CITATION = JMAPA,12,498;%%
\bibitem [{\citenamefont {Lovelock}(1972)}]{1972-Lovelock-JMP}%
  \BibitemOpen
  \bibfield  {author} {\bibinfo {author} {\bibfnamefont {D.}~\bibnamefont
  {Lovelock}},\ }\href {\doibase 10.1063/1.1666069} {\bibfield  {journal}
  {\bibinfo  {journal} {J. Math. Phys.}\ }\textbf {\bibinfo {volume} {13}},\
  \bibinfo {pages} {874} (\bibinfo {year} {1972})}\BibitemShut {NoStop}%
%%CITATION = JMAPA,13,874;%%
\bibitem [{\citenamefont {{Zumino}}(1986)}]{1986-Zumino-PRep}%
  \BibitemOpen
  \bibfield  {author} {\bibinfo {author} {\bibfnamefont {B.}~\bibnamefont
  {{Zumino}}},\ }\href {\doibase 10.1016/0370-1573(86)90076-1} {\bibfield
  {journal} {\bibinfo  {journal} {Phys. Rept.}\ }\textbf {\bibinfo {volume}
  {137}},\ \bibinfo {pages} {109} (\bibinfo {year} {1986})}\BibitemShut
  {NoStop}%
\bibitem [{\citenamefont {Padmanabhan}\ and\ \citenamefont
  {Kothawala}(2013)}]{2013-Padmanabhan.Kothawala-PRep}%
  \BibitemOpen
  \bibfield  {author} {\bibinfo {author} {\bibfnamefont {T.}~\bibnamefont
  {Padmanabhan}}\ and\ \bibinfo {author} {\bibfnamefont {D.}~\bibnamefont
  {Kothawala}},\ }\href {\doibase 10.1016/j.physrep.2013.05.007} {\bibfield
  {journal} {\bibinfo  {journal} {Phys. Rept.}\ }\textbf {\bibinfo {volume}
  {531}},\ \bibinfo {pages} {115} (\bibinfo {year} {2013})},\ \Eprint
  {http://arxiv.org/abs/1302.2151} {arXiv:1302.2151 [gr-qc]} \BibitemShut
  {NoStop}%
%%CITATION = ARXIV:1302.2151;%%
\bibitem [{\citenamefont {Nandi}\ and\ \citenamefont
  {Shankaranarayanan}(2018)}]{2017-Nandi.Shankaranarayanan-JCAP}%
  \BibitemOpen
  \bibfield  {author} {\bibinfo {author} {\bibfnamefont {D.}~\bibnamefont
  {Nandi}}\ and\ \bibinfo {author} {\bibfnamefont {S.}~\bibnamefont
  {Shankaranarayanan}},\ }\href {\doibase 10.1088/1475-7516/2018/01/039}
  {\bibfield  {journal} {\bibinfo  {journal} {JCAP}\ }\textbf {\bibinfo
  {volume} {1801}},\ \bibinfo {pages} {039} (\bibinfo {year} {2018})},\ \Eprint
  {http://arxiv.org/abs/1704.06897} {arXiv:1704.06897 [astro-ph.CO]}
  \BibitemShut {NoStop}%
%%CITATION = ARXIV:1704.06897;%%
\bibitem [{\citenamefont {Heisenberg}\ \emph {et~al.}(2018)\citenamefont
  {Heisenberg}, \citenamefont {Kase},\ and\ \citenamefont
  {Tsujikawa}}]{2018-Heisenberg.etal-PRD}%
  \BibitemOpen
  \bibfield  {author} {\bibinfo {author} {\bibfnamefont {L.}~\bibnamefont
  {Heisenberg}}, \bibinfo {author} {\bibfnamefont {R.}~\bibnamefont {Kase}}, \
  and\ \bibinfo {author} {\bibfnamefont {S.}~\bibnamefont {Tsujikawa}},\ }\href
  {\doibase 10.1103/PhysRevD.98.024038} {\bibfield  {journal} {\bibinfo
  {journal} {Phys. Rev.}\ }\textbf {\bibinfo {volume} {D98}},\ \bibinfo {pages}
  {024038} (\bibinfo {year} {2018})},\ \Eprint
  {http://arxiv.org/abs/1805.01066} {arXiv:1805.01066} \BibitemShut {NoStop}%
%%CITATION = ARXIV:1805.01066;%%
\bibitem [{\citenamefont {Nicolis}\ \emph {et~al.}(2009)\citenamefont
  {Nicolis}, \citenamefont {Rattazzi},\ and\ \citenamefont
  {Trincherini}}]{2009-Nicolis-PRD}%
  \BibitemOpen
  \bibfield  {author} {\bibinfo {author} {\bibfnamefont {A.}~\bibnamefont
  {Nicolis}}, \bibinfo {author} {\bibfnamefont {R.}~\bibnamefont {Rattazzi}}, \
  and\ \bibinfo {author} {\bibfnamefont {E.}~\bibnamefont {Trincherini}},\
  }\href {\doibase 10.1103/PhysRevD.79.064036} {\bibfield  {journal} {\bibinfo
  {journal} {Phys. Rev.}\ }\textbf {\bibinfo {volume} {D79}},\ \bibinfo {pages}
  {064036} (\bibinfo {year} {2009})},\ \Eprint {http://arxiv.org/abs/0811.2197}
  {arXiv:0811.2197 [hep-th]} \BibitemShut {NoStop}%
%%CITATION = ARXIV:0811.2197;%%
\bibitem [{\citenamefont {Deffayet}\ \emph
  {et~al.}(2009{\natexlab{b}})\citenamefont {Deffayet}, \citenamefont
  {Esposito-Farese},\ and\ \citenamefont {Vikman}}]{Deffayet2009}%
  \BibitemOpen
  \bibfield  {author} {\bibinfo {author} {\bibfnamefont {C.}~\bibnamefont
  {Deffayet}}, \bibinfo {author} {\bibfnamefont {G.}~\bibnamefont
  {Esposito-Farese}}, \ and\ \bibinfo {author} {\bibfnamefont {A.}~\bibnamefont
  {Vikman}},\ }\href {\doibase 10.1103/PhysRevD.79.084003} {\bibfield
  {journal} {\bibinfo  {journal} {Phys. Rev.}\ }\textbf {\bibinfo {volume}
  {D79}},\ \bibinfo {pages} {084003} (\bibinfo {year} {2009}{\natexlab{b}})},\
  \Eprint {http://arxiv.org/abs/0901.1314} {arXiv:0901.1314 [hep-th]}
  \BibitemShut {NoStop}%
%%CITATION = ARXIV:0901.1314;%%
\bibitem [{\citenamefont {Goon}\ \emph {et~al.}(2012)\citenamefont {Goon},
  \citenamefont {Hinterbichler}, \citenamefont {Joyce},\ and\ \citenamefont
  {Trodden}}]{Goon:2012mu}%
  \BibitemOpen
  \bibfield  {author} {\bibinfo {author} {\bibfnamefont {G.}~\bibnamefont
  {Goon}}, \bibinfo {author} {\bibfnamefont {K.}~\bibnamefont {Hinterbichler}},
  \bibinfo {author} {\bibfnamefont {A.}~\bibnamefont {Joyce}}, \ and\ \bibinfo
  {author} {\bibfnamefont {M.}~\bibnamefont {Trodden}},\ }\href {\doibase
  10.1016/j.physletb.2012.06.065} {\bibfield  {journal} {\bibinfo  {journal}
  {Phys. Lett.}\ }\textbf {\bibinfo {volume} {B714}},\ \bibinfo {pages} {115}
  (\bibinfo {year} {2012})},\ \Eprint {http://arxiv.org/abs/1201.0015}
  {arXiv:1201.0015 [hep-th]} \BibitemShut {NoStop}%
%%CITATION = ARXIV:1201.0015;%%
\bibitem [{\citenamefont {Peeters}(2007{\natexlab{a}})}]{2007-Peeters1-arXiv}%
  \BibitemOpen
  \bibfield  {author} {\bibinfo {author} {\bibfnamefont {K.}~\bibnamefont
  {Peeters}},\ }\href {\doibase 10.1016/j.cpc.2007.01.003} {\bibfield
  {journal} {\bibinfo  {journal} {Comput. Phys. Commun.}\ }\textbf {\bibinfo
  {volume} {176}},\ \bibinfo {pages} {550} (\bibinfo {year}
  {2007}{\natexlab{a}})},\ \Eprint {http://arxiv.org/abs/cs/0608005}
  {arXiv:cs/0608005 [cs.SC]} \BibitemShut {NoStop}%
%%CITATION = CS/0608005;%%
\bibitem [{\citenamefont {Peeters}(2007{\natexlab{b}})}]{2007-Peeters2-arXiv}%
  \BibitemOpen
  \bibfield  {author} {\bibinfo {author} {\bibfnamefont {K.}~\bibnamefont
  {Peeters}},\ }\href@noop {} {\  (\bibinfo {year} {2007}{\natexlab{b}})},\
  \Eprint {http://arxiv.org/abs/hep-th/0701238} {arXiv:hep-th/0701238 [hep-th]}
  \BibitemShut {NoStop}%
%%CITATION = HEP-TH/0701238;%%
\end{thebibliography}
\end{document}